\documentclass[aps,prc,twocolumn,showpacs,floatfix,preprintnumbers,superscriptaddress,amsmath,amssymb,longbibliography]{revtex4-1}

\usepackage{pifont}
\usepackage[colorlinks,citecolor=blue]{hyperref}
\usepackage[colorinlistoftodos]{todonotes}
\usepackage[normalem]{ulem}
\usepackage{color}
\usepackage[utf8]{inputenc}
\usepackage{bm}

\makeatletter
\def\@bibdataout@aps{%
\immediate\write\@bibdataout{%
@CONTROL{%
apsrev41Control%
\longbibliography@sw{%
    ,author="08",editor="1",pages="1",title="0",year="1"%
    }{%
    ,author="08",editor="1",pages="1",title="",year="1"%
    }%
  }%
}%
\if@filesw \immediate \write \@auxout {\string \citation {apsrev41Control}}\fi 
}
\makeatother

\graphicspath{{./}{figures/}}

\definecolor{pastelgray}{rgb}{0.81, 0.81, 0.77}
\definecolor{beaublue}{rgb}{0.9, 0.9, 0.93}

\newcommand{\figc}[1]{{Fig.~\ref{#1}}}

\newcommand{\rpa}{\textit{r}-process}
\newcommand{\cmark}{{\color{teal}\ding{51}}}
\newcommand{\xmark}{{\color{magenta}\ding{55}}}

\begin{document}
\title{Theoretical description  of fission yields: towards a fast and efficient global model}

\author{Jhilam Sadhukhan}
\affiliation{Physics Group, Variable Energy Cyclotron Centre, Kolkata 700064, India }
\affiliation{Homi Bhabha National Institute, Mumbai 400094, India }

\author{Samuel A. Giuliani}
\affiliation{FRIB Laboratory, Michigan State University, East Lansing, Michigan 48824, USA}
\affiliation{European Centre for Theoretical Studies in Nuclear Physics and Related Areas (ECT*-FBK), Trento, Italy.}
\affiliation{Department of Physics, Faculty of Engineering and Physical Sciences, University of Surrey, Guildford, Surrey GU2 7XH, United Kingdom}

\author{Witold Nazarewicz}
\affiliation{FRIB Laboratory and Department of Physics and Astronomy,  Michigan State University, East Lansing, Michigan 48824, USA}

\date{\today}

\begin{abstract}
\begin{description}
\item[Background] A quantitative microscopic understanding of the fission-fragment yield distributions represents a major challenge for nuclear theory as it involves the intricate competition between large-amplitude nuclear collective motion and single-particle nucleonic motion. 
\item[Purpose] A recently proposed approach to global modeling of fission fragment distributions is extended to account for odd-even staggering in charge yields and for neutron evaporation. 
\item[Method]Fission trajectories  are obtained within the density functional theory framework, allowing for a microscopic determination of the most probable fission prefragment configurations. Mass and charge yields distributions are  constructed by means of a statistical approach rooted in a microcanonical ensemble.
\item[Result]We show that the proposed hybrid model can reproduce experimental mass and charge fragment yields, including the odd-even staggering,  for a wide range of fissioning nuclei.  Experimental isotopic yields can be  described within a simple neutron evaporation scheme. We also explore fission fragment distributions of exotic neutron-rich and superheavy  systems, and compare our predictions with other state-of-the art global calculations.
\item[Conclusion] Our study suggests that the microscopic rearrangement of nucleons into fission fragments occurs well before the scission, and that the subsequent dynamics is mainly driven by the thermal excitations and bulk features of the nuclear binding. The proposed simple hybrid approach is  well suited for large-scale calculations
involving hundreds of fissioning nuclei.
\end{description}
\end{abstract}
\maketitle

\section{Introduction}
Fission is a fundamental nuclear decay that plays a crucial role in many areas of science, ranging from the design of nuclear reactors to studies devoted to physics beyond the standard model of particle physics~\cite{vogel2015}, and the synthesis of heavy elements~\cite{horowitz2018,Cowan2021}. This process is driven by both the nuclear large-amplitude collective motion and the quantum mechanical shell effects rooted in the single-particle motion of individual nucleons. The yield patterns of  fission fragments involve an intricate interplay between shell structure and  pairing correlations associated with nuclear superfluidity, and stochastic effects. Therefore, a predictive microscopic description of this complex process constitutes a great  challenge for  nuclear theory~\cite{bender2020}. In particular, current global models applied to systematic studies of fission fragments distributions cannot consistently explain the observed enhanced production of fragments~\cite{amiel1975} together with other fission-yield characteristics such as  distribution peaks and widths.  

Within the fission realm, odd-even staggering (OES) in charge distributions has been traditionally attributed to the dissipative coupling between the collective and  individual (or intrinsic) degrees of freedom. During the descent towards scission, nucleonic Cooper pairs can be broken by absorbing the intrinsic excitation energy produced via the dissipation of the collective kinetic energy. Signature of this pair-breaking mechanism has been observed in the experimental data on average kinetic energy of fragments in low energy fission~\cite{lang1980,mariolopoulos1981}. However, this picture has been challenged by measurements showing a correlation between the OES in charge yields and the mass asymmetry of the fission fragments~\cite{caamano2011}, a phenomenon unrelated to energy dissipation.

From a microscopic point of view, OES in charge yields can be related to the dynamical breaking of Cooper pairs in avoided-level-crossing regions, where the  Landau-Zener effect  results in  low-lying time dependent excitations~\cite{mirea2014,Mirea2017}. In an apparently uncorrelated manner, peak positions in fragment distributions are mainly governed by the shell effects determining the most probable fission configuration, which, within certain models, may be manifested through the topology of the collective potential energy surface (PES). In contrast, the distribution width  is primarily driven by stochastic effects that allow the population of highly mass-asymmetric fission configurations~\cite{sadhukhan2017,sierk2017}.

Although the underlying mechanisms are  established qualitatively, current state-of-the-art theoretical models struggle to obtain a coherent quantitative description of the gross characteristics of fragment distributions and the OES in charge yields~\cite{bender2020}. For example,  scission-point models (SPM)~\cite{lemaitre2019,Carjan2019,Pasca2019} take into account the statistical distributions~\cite{Dec68,Dem01} required to reproduce OES, but the resulting mass distributions lack dynamical correlations. The latter are considered by  models employing  the Brownian shape-motion approach  (BSM) in a multidimensional configuration space~\cite{Mumpower2020,Albertsson2020,ishizuka2017,sadhukhan2016} that can  take into account the dynamical effects such as  dissipation and configuration changes during the descent towards scission. However, the interplay between the dynamics and the thermalization process is yet to be explored in a comprehensive manner \cite{bender2020,Bulgac2019a}. 

While some attempts have been made to include OES within the BSM formalism~\cite{moller2015c,Verriere2021b}, the Brownian dynamics on macroscopic-microscopic PESs overestimates~\cite{Mumpower2020} the widths of fission yields for very heavy systems. Predictions of fission-fragment yields~\cite{goddard2015,tanimura2017} and characteristics of fission pathways~\cite{scamps2018a,scamps2019a} in selected nuclei have been also obtained using self-consistent time-dependent approaches such as the time-dependent density functional theory (DFT) 
or the time-dependent generator coordinate method (TDGCM)~\cite{regnier2019,zhao2020,verriere2020}. Such calculations properly account for shell effects but have  limitations when it comes to the treatment of dissipative dynamics and associated fluctuations. Even though within these models the OES could be obtained by means of the particle-number projection~\cite{Simenel2010,scamps2015,Verriere2019}, recent TDGCM calculations illustrate  the difficulties  in reproducing the experimental OES in charge yields~\cite{Verriere2021}.

In this study, we predict nuclear fission-fragment yield distributions using an extension of a recently developed  framework~\cite{Sadhukhan2020} that combines microscopic input obtained with nuclear DFT with a simple statistical model rooted in a  microcanonical ensemble. We demonstrate that such a hybrid approach, well-suited for large-scale calculations involving hundreds of nuclei, can consistently explain the experimental data for a wide range of fissioning nuclei and make predictions for unknown systems.  

\section{Formalism}\label{sec:formalism}
\begin{figure}[tb]
	\includegraphics[width=0.8\columnwidth]{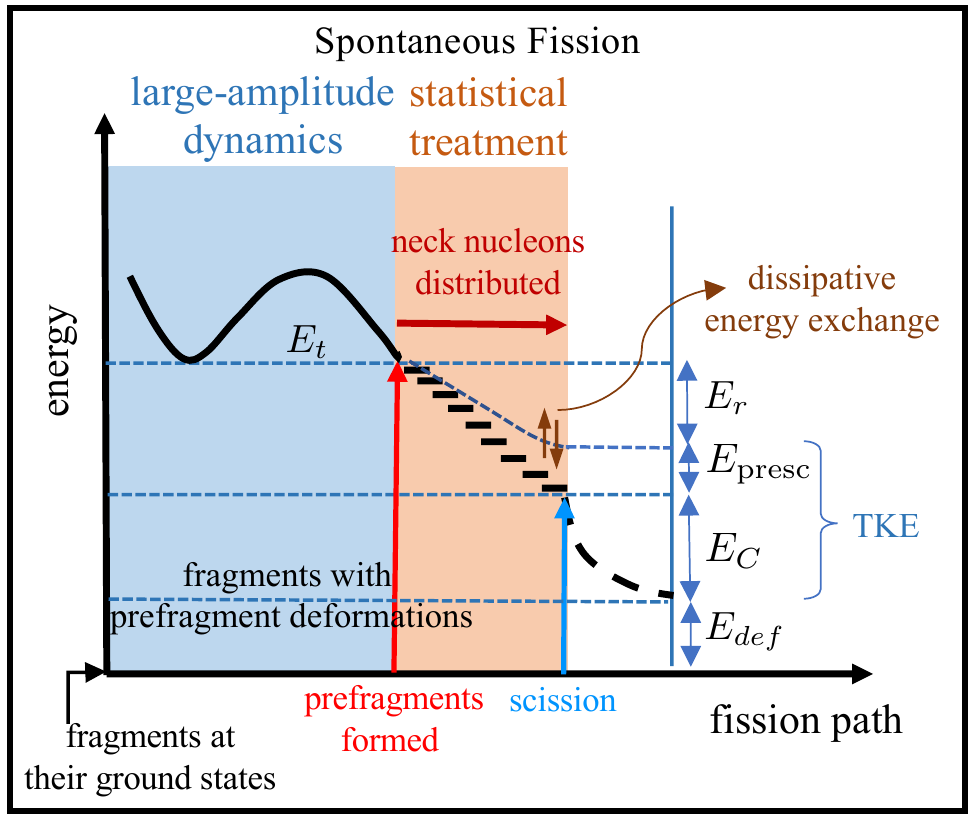}
	\caption{Schematic representation of energy sharing and their evolution along the fission pathway for spontaneous fission (adapted from Ref.~\cite{Caamano2017}). In thermal fission, the excess energy ($\sim 6$~MeV) is shared among  the  pre-scission  collective  kinetic  energy  of  the fragments $E_\textrm{presc}$ and the residual thermal energy $E_r$. $E_C$ and $E_{def}$ are the Coulomb and deformation energies, respectively, and TKE is the total kinetic energy of the fragments. See text for details.}
	\label{fig:schematic}
\end{figure}

Fission fragment distributions are obtained in this study by extending the methodology described in Ref.~\cite{Sadhukhan2020}, which we briefly summarize in the following.  We recall that contrary to most of the approaches to fission-fragment yields, which rely on the choice of near-scission configurations, the predictions of our model  are based on pre-scission configurations, which are  less sensitive to the dimension of the collective space. Moreover, as nucleonic localizations suggest~\cite{zhang2016,sadhukhan2017}, apart from the neck region, microscopic arrangement of nucleons in prefragments  quickly stabilizes after reaching the pre-scission configuration. Therefore, the subsequent dynamics in the configuration space may become less critical for deciding the population of different fragments, and a redistribution of neck nucleons based on statistical phase-space arguments seems to be sufficient to determine the final fragment distribution~\cite{Sadhukhan2020}. The complete energy evolution of a fissioning system along the primary fission degree of freedom is shown schematically in Fig.~\ref{fig:schematic}.  

The first step in the estimation of the fission yields is the calculation of the PES, which in our study is obtained  by constraining  the mass quadrupole moment $Q_{20}$ and the mass octupole moment $Q_{30}$. These two collective coordinates are sufficient for the fission-yield identification~\cite{matheson2019}. For spontaneous fission (SF), the PES is computed at zero temperature while for induced fission it is obtained by constraining the temperature to the excitation energy of the compound system. To this end, we use the finite-temperature  approach of Ref.~\cite{pei2009} and solve the temperature-dependent Hartree-Fock-Bogoliubov equations by employing the DFT solver HFODD~\cite{schunck2017}. The weak sensitivity of yields distributions  to the choice of energy density functionals  has been established in our previous work~\cite{Sadhukhan2020}. Here we  consider the Skyrme parametrization SkM$^\ast$~\cite{bartel1982} in the particle-hole channel. In the pairing channel, we take the mixed-type density-dependent delta interaction~\cite{dobaczewski2002}.

For SF, the most probable fission path is obtained by minimizing the collective action integral in a two-dimensional collective space. The action integral can be written as
\begin{equation}\label{eq:Sl}
	S(L) = \frac{1}{\hbar^{2}} \int _{s_\textrm{in}} ^{s_\textrm{out}}
	\sqrt{2 \mathcal{M}_\textrm{eff}(s) \left[V(s) - E_0\right]} \, ds \,,
\end{equation}
where $\mathcal{M}_\textrm{eff}(s)$ is the collective inertia, $V(s)$ the potential energy, and $E_0$ represents the collective ground-state energy. Here, $ds$ is the element of length along the collective path $L(s)$ with $s_\textrm{in}$ and $s_\textrm{out}$ being the inner and outer turning points, respectively. Although the SF half-life is very sensitive to the choice of  $\mathcal{M}_\textrm{eff}$ and $E_0$, a constant $\mathcal{M}_\textrm{eff}$ given by its ground-state value works reasonably well for the present purpose~(see Fig.~3(d) in \cite{Sadhukhan2020}) and any value of $E_0$ in the range of 0--1 MeV hardly affects the configuration at $s_\textrm{out}$, which is selected as the pre-scission configuration for subsequent calculations. 

Induced fission can be viewed as a diffusive process taking place above the collective potential barrier.  In this case, we extract the minimum-potential path by joining the local minima starting from the lowest-energy configuration. On this path, we identify the pre-scission configuration by matching the potential energy outside the barrier region with that of the lowest-energy configuration. 

Once the pre-scission configuration has been found, the proton and neutron numbers of the fission {\it prefragments} and their deformations  for this configuration are identified by means of the nucleon localization function (NLF)~\cite{zhang2016,sadhukhan2017}. The remaining nucleons, which belong to the neck connecting the prefragments, are then distributed among the two prefragments according to the microcanonical probability describing the population of the final fragments~\cite{fong1953,Bondorf1995}:
\begin{eqnarray}\label{eq:prob}
	&P(A_1,A_2) \propto \sqrt{ \left(\frac{(A_1A_2)^8}{\left(A_1^{5/3}+A_2^{5/3}\right)^3(A_1+A_2)^3}\right) \frac{a_1 a_2}{(a_1+a_2)^5} 
	}
	\nonumber\\
	&\times\left(1-\frac{1}{2\sqrt{(a_1+a_2)E_r}}\right)
	E_r^{9/4}\exp{\left\{2\sqrt{(a_1+a_2)E_r}\right\}},
\end{eqnarray}
where  $a_i=A_i/10$~MeV$^{-1}$ is the level density parameter of fragment
$(A_i,Z_i)$, with $i=1$ or 2, or L(ight) or H(eavy). Modified forms of Eq.~(\ref{eq:prob}) have also been proposed~\cite{Fong56}, but the resulting fragment distributions are found to be practically indistinguishable from the results presented in this work. As shown in  Fig.~\ref{fig:schematic}, the residual thermal energy of each fragment combination is
\begin{equation}
	E_r = E_t - \left\{E_b^{\rm L} + E_b^{\rm H} +E_{C}+E_\textrm{presc}\right\}.
\label{eq:enrg}
\end{equation} 
In the above expression,  $E_t$ is the energy of the fissioning nucleus extracted from the pre-scission configuration, $E_{C}(\beta^{\rm L},\beta^{\rm H})$ is the electrostatic repulsion between the fragments characterized by deformations $\beta^{\rm L},\beta^{\rm H}$, and  $E_b^i(\beta^i)$ is the binding energy of the fragment $i$, having deformation $\beta^i$, estimated from the deformed liquid drop model (LDM)~\cite{myers1966} that properly describes the bulk properties of a nucleus. We shall point out that we explicitly avoid the microscopic calculations of $E^i_b(\beta_i)$ in order to preserve the most favored configurations predicted by the microscopically-defined prefragments which are driven by combined effort of deformed shell effects and collective dynamics. The importance of this choice is discussed in Sec.~\ref{sec:result}. In this respect, we emphasize that our model is distinct from SPM where accurate binding energies of the fragments at their ground states  are used to estimate $E_r$. In our approach, we employ Eq.~(\ref{eq:prob}) to get the distribution of neck nucleons comprising of only $\sim$10\%~\cite{Sadhukhan2020} of the total nuclear mass. The configuration space of our model is thus given by all the possible combination of the neck nucleons complementing the two prefragments. 

As shown in Fig.~\ref{fig:schematic}, fragment deformations at scission are assumed to be the same as prefragment deformations, since prefragments contain the majority of nucleons that are stabilized by shell effects. Further, according to the construction of prefragments, axially symmetric quadrupole deformation ($\beta\equiv\beta_2$) is most important. Consequently, we incorporated quadrupole deformations of light  and heavy fragments in $E_C(\beta_2^{\rm L},\beta_2^{\rm H})$ as given in~\cite{won73}, Eq.~(13). The Coulomb and surface terms in $E^i_{b}(\beta_2^i)$ are also corrected for shape deformations $\beta_2^i$. We found that $\beta_2^{\rm H,L}\lesssim0.1$ for all the prefragments considered in the present work and in~\cite{Sadhukhan2020}. Further, as demonstrated in Sec.~\ref{sec:result}, yield distributions are insensitive to prefragment deformations for such small values of $\beta_2^{\rm H,L}$ and, therefore, one can safely assume $\beta_2^{\rm L}=\beta_2^{\rm H}=0$. Nevertheless, our model can take care of non-zero $\beta_2^{\rm H,L}$.

The term $E_\textrm{presc}$ in Eq.~\eqref{eq:enrg} represents the pre-scission collective kinetic energy of the fragments. In low-energy fission, this quantity can vary in a range of 0--20~MeV~\cite{Caamano2017} resulting an uncertainty in $E_r$. Furthermore, within an ensemble, $E_r$ for different events fluctuate by $\sim10$~MeV~\cite{Caamano2017} due to dissipative energy transfer from $E_\textrm{presc}$. Finally, since $E_t$ and $E_b^i$ in Eq.~(\ref{eq:enrg}) are obtained from two different prescriptions, $E_r$ is required to be scaled such as the upper limit matches the typical maximum excitation energy of fission fragments $E_r^\text{max} = 40$\,MeV~\cite{Caamano2017}, measured in low-energy fission. As shown in Sec.~\ref{sec:result}, we find that fission and charge yields predicted by our model are virtually insensitive to $E_r$ within a large range of values.

In order to account for odd-even effects, we augmented the LDM expression by the smooth pairing term of Ref.~\cite{bertsch2009}:
\begin{equation}
	\tilde{\Delta} = \frac{c}{A_i^{\alpha}} \,,
\label{eq:pair}
\end{equation}
with $\alpha=0.31$ and $c=4.66~(4.31)$ for neutron (protons). This term increases the binding energy of even-even nuclei with respect to their odd-$A$ neighbors. 

Subsequently, we consider the  neutron evaporation from primary fragments. A microscopic handling of the neutron multiplicity of fission fragments  is a daunting task~\cite{bender2020}. First, it is unclear whether the excitation energy partitioning occurs in a condition of thermal equilibrium. Second, the deexcitation of the nascent fragments is driven by the competition between different decay channels including neutron emission, electromagnetic radiation, and beta decay. In the present work, we simply assume that the neutron emission is statistical in nature, and that the total excitation energy of the fragments is distributed uniformly among all degrees of freedom. In our simulations, each excited fragment is allowed to emit neutrons until its excitation energy falls below the neutron emission threshold given by the neutron  separation energy $S_n$. After each neutron emission, the excitation energy is adjusted to $E_r^{\prime}=E_r-S_n-E_n$, where $E_n$ is the average kinetic energy of the emitted neutron.

Following the standard procedure~\cite{sadhukhan2016,regnier2016}, the  mass distributions are convoluted using a Gaussian smoothing function with a width $\sigma=3$. For charge distributions, odd and even atomic numbers are first convoluted separately with a Gaussian function ($\sigma=2$), and then the full distribution is  convoluted with another Gaussian function ($\sigma=0.5$). This procedure preserves the OES while reproducing the width of experimental charge distributions. We wish to emphasize that all parameters of our model are fixed globally, i.e., no adjustments are needed when making predictions for individual nuclei.

In Table~\ref{tab:models_comp}  we briefly compare the basic features of our model with those of SPM and BSM approaches that have been employed in large-scale systematic calculations of fission-fragment distributions. While several implementations of these  frameworks exist, the benchmark results presented in this paper correspond to the recent  state-of-the-art global calculations: a modified version of the Scission-Point Yield 2 (SPY2) model~\cite{lemaitre2021} and the BSM model  of Ref.~\cite{Mumpower2020}. (The BSM predictions  for the OES and the total kinetic energy  are not included in the survey of Ref.~\cite{Mumpower2020}.)

\begin{table}[tb]
       \caption{Comparison between global models of fission-fragment distributions. }
        \small
\begin{ruledtabular}
	\begin{tabular}{lccc}
	Feature	& This work	& ~~BSM~~	& ~~SPM~~   \\
	\hline
	Odd-even staggering & \cmark	& \cmark 	& \cmark \\
	Dynamics                 	& \cmark	& \cmark	& \xmark \\
	Microscopic PES	& \cmark	& \xmark	& \cmark \\
	Total kinetic energy 		& \xmark	& \cmark  	& \cmark \\
	Spontaneous fission & \cmark	& \xmark	& \cmark \\
	Induced fission & \cmark	& \cmark	& \cmark \\
	Scission config. essential 	& \xmark	& \cmark	& \cmark\\
	\end{tabular}
	\label{tab:models_comp}
	\end{ruledtabular}
\end{table}

%
%
\section{Results}\label{sec:result}
\subsection{Model validation and sensitivity tests}
\begin{figure}[tbh]
	\centering\includegraphics[width=\columnwidth]{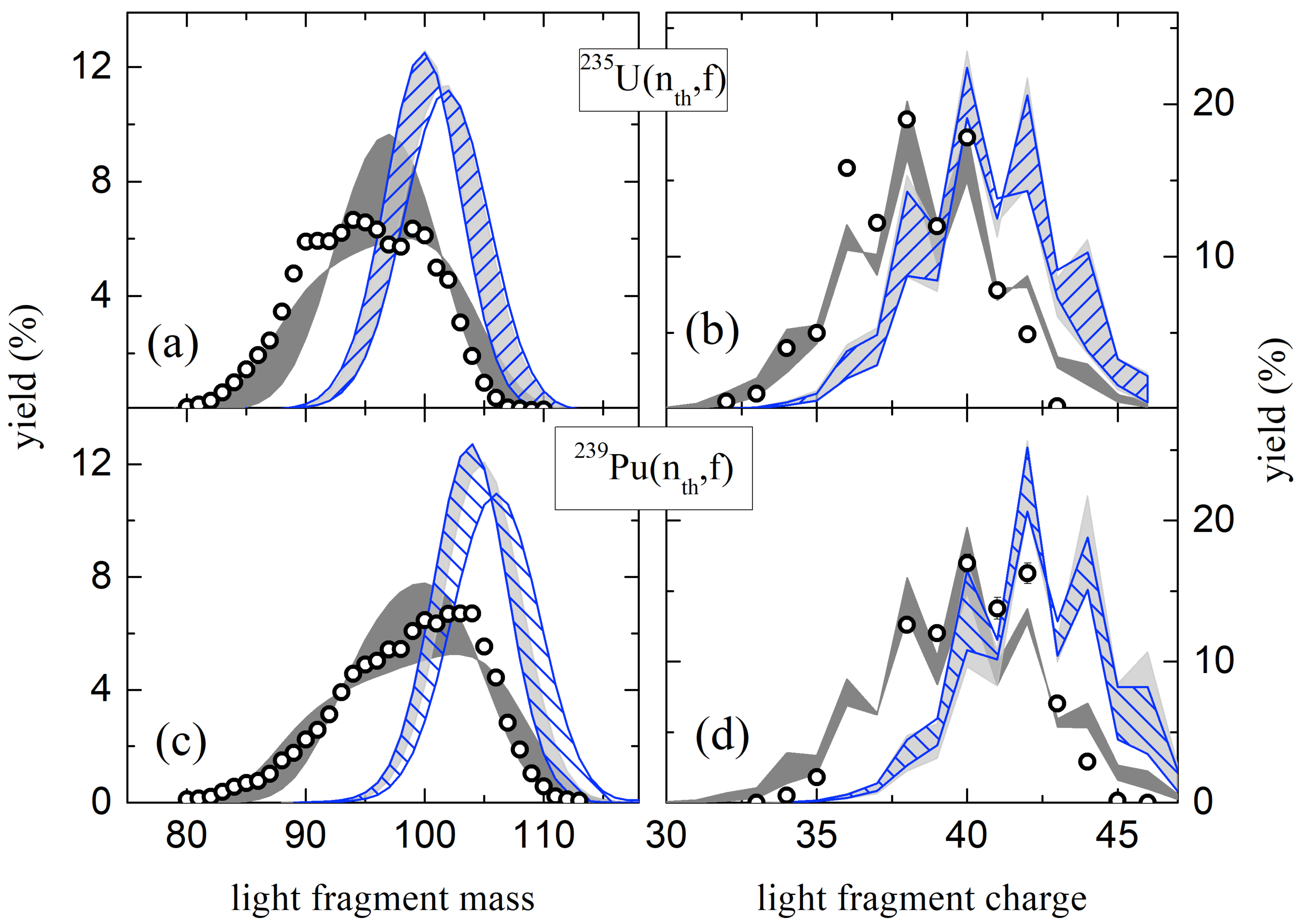}
	\caption{Calculated mass (left panels) and charge (right panels) distributions of fission fragments using the model described in this study (dark bands), the SkM$^*$ mass table (light band), and AME2020 experimental masses (blue pattern). Experimental yields are marked by circles~\cite{lang1980,laidler1962}. Widths in the calculated results come from two-particle uncertainty~\cite{Sadhukhan2020}.}
	\label{fig:BEtest}
\end{figure}

We first justify our proposition on the use of LDM in estimating the fragment binding energies $E^i_b$. Fragment yields of two well-known fission reactions are shown in Fig.~\ref{fig:BEtest}, where ground-state binding energies of the fragments are used in Eq.~\eqref{eq:enrg} instead of the LDM values. Calculations are performed for two different mass tables: the SkM$^*$ mass table~\cite{massexplorer} and the 
experimental AME2020 atomic mass evaluation~\cite{AME20}. Both mass tables result in an OES in charge yields but   both variants underestimate the measured mass asymmetries and widths of the mass distributions. This is due to the presence of ground-state shell effects in the mass tables, which gives rise to the overestimation of the shell effects. As already mentioned, the shell structure of the fragments is decided dynamically in the prefragments, restricting the configuration space in the  statistical treatment. Therefore, ground-state shell corrections counteract the deformed shell effects driving the prefragment localization and shift the peak's location towards more symmetric configurations. From this observation, we conclude that our prescription can be viewed as a phenomenological ansatz that has been justified {\it a posteriori}.

\begin{figure}[tb]
	\centering\includegraphics[width=\columnwidth]{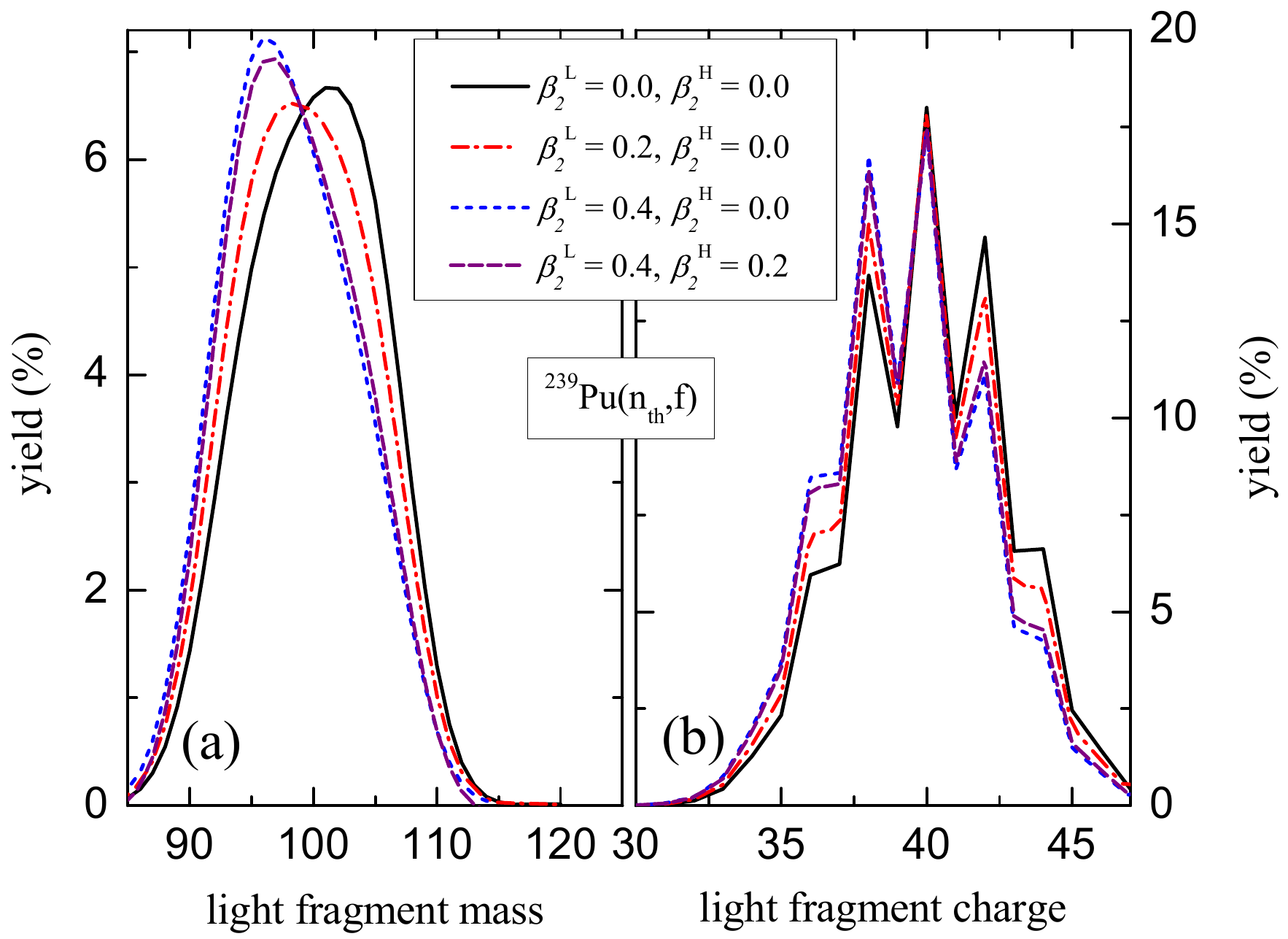}
	\caption{$^{239} \textrm{Pu} (n_\textrm{th},f)$ mass (left panels) and charge (right panels) fission fragment distributions calculated using the model described in this study for different values of heavy fragment ($\beta_2^H$) and light fragment ($\beta_2^L$) deformations. These are obtained without two-particle uncertainties.}
	\label{fig:defortest}
\end{figure}

As mentioned in Sec.~\ref{sec:formalism}, we found that $\beta_2^i$ for all the prefragments calculated in this work are small. Nevertheless, to test the robustness of our model, we plot the yield distributions in Fig.~\ref{fig:defortest} for a wide variation of $\beta_2^{\rm L}$. Here we consider lower values for $\beta_2^{\rm H}$ as the heavy prefragment is usually close to the doubly-magic $^{132}$Sn for the chosen nucleus ~\cite{sadhukhan2017}. Evidently, corrections to yields due to shape deformations are small compared to the two-particle uncertainty defined in~\cite{Sadhukhan2020} and also shown in Fig.~\ref{fig:BEtest} for the same system.

\begin{figure}[tb]
	\centering\includegraphics[width=\columnwidth]{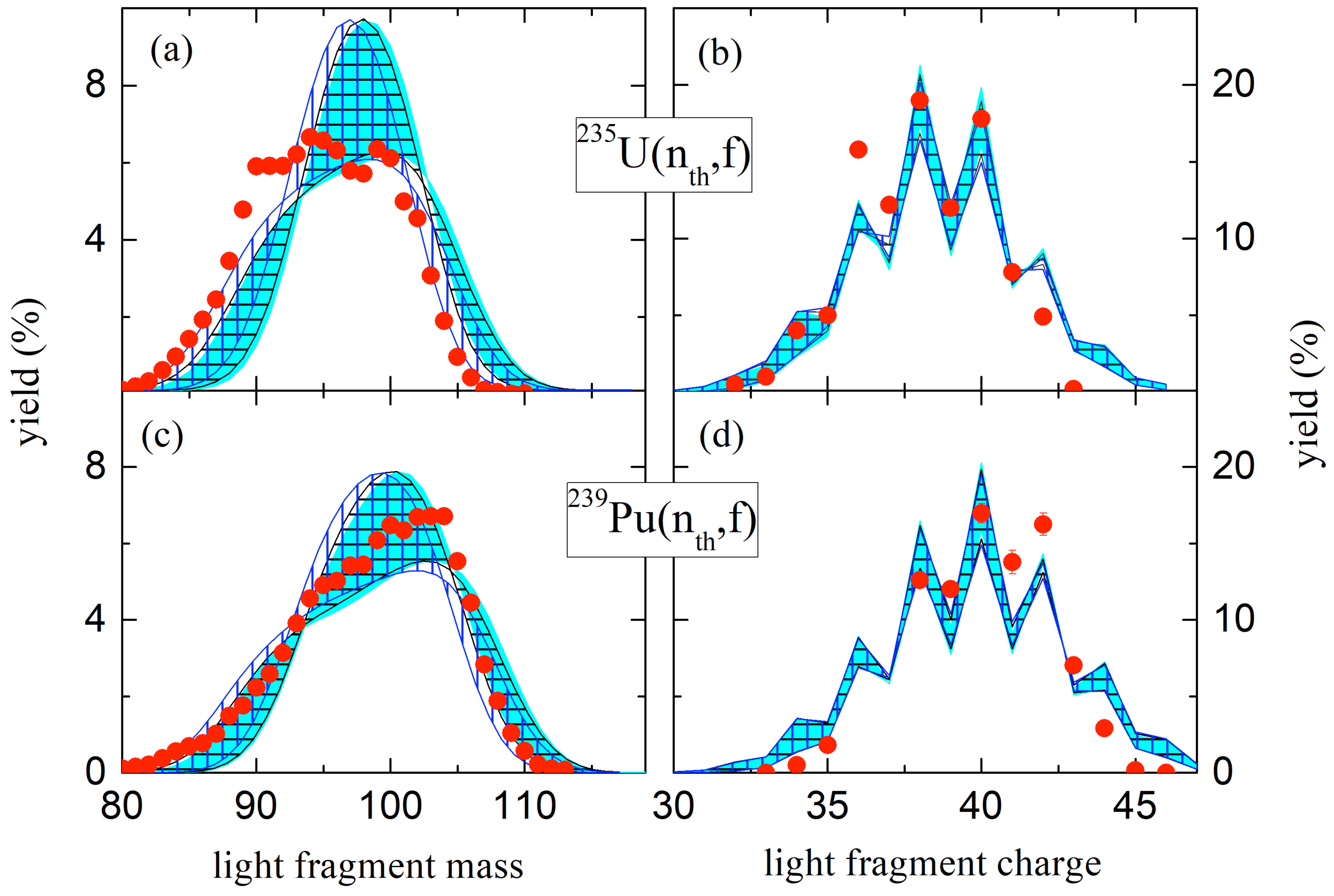}
	\caption{Calculated mass (left panels) and charge (right panels) fission fragment distributions using our model for $E_r^{\text{max}}=20$\,MeV (solid blue band), 30\,MeV (horizontal pattern), and 40\,MeV (vertical pattern). Experimental yields are marked by  circles~\cite{lang1980,laidler1962}.}
	\label{fig:Etest}
\end{figure}

The sensitivity of our results to $E_r^\text{max}$ is presented in Fig.~\ref{fig:Etest}, where yield distributions are calculated for three different values of $E_r^{\text{max}}$ covering a broad range of possible residual energy. The charge yield distributions are found to be indistinguishable for this range of $E_r^{\text{max}}$, and mass yields shift marginally toward lower masses due to neutron evaporation (see also discussion below). This demonstrates that, within a reasonable range, our results are  insensitive to $E_r^{\text{max}}$.
\begin{figure}[tb]
	\centering\includegraphics[width=\columnwidth]{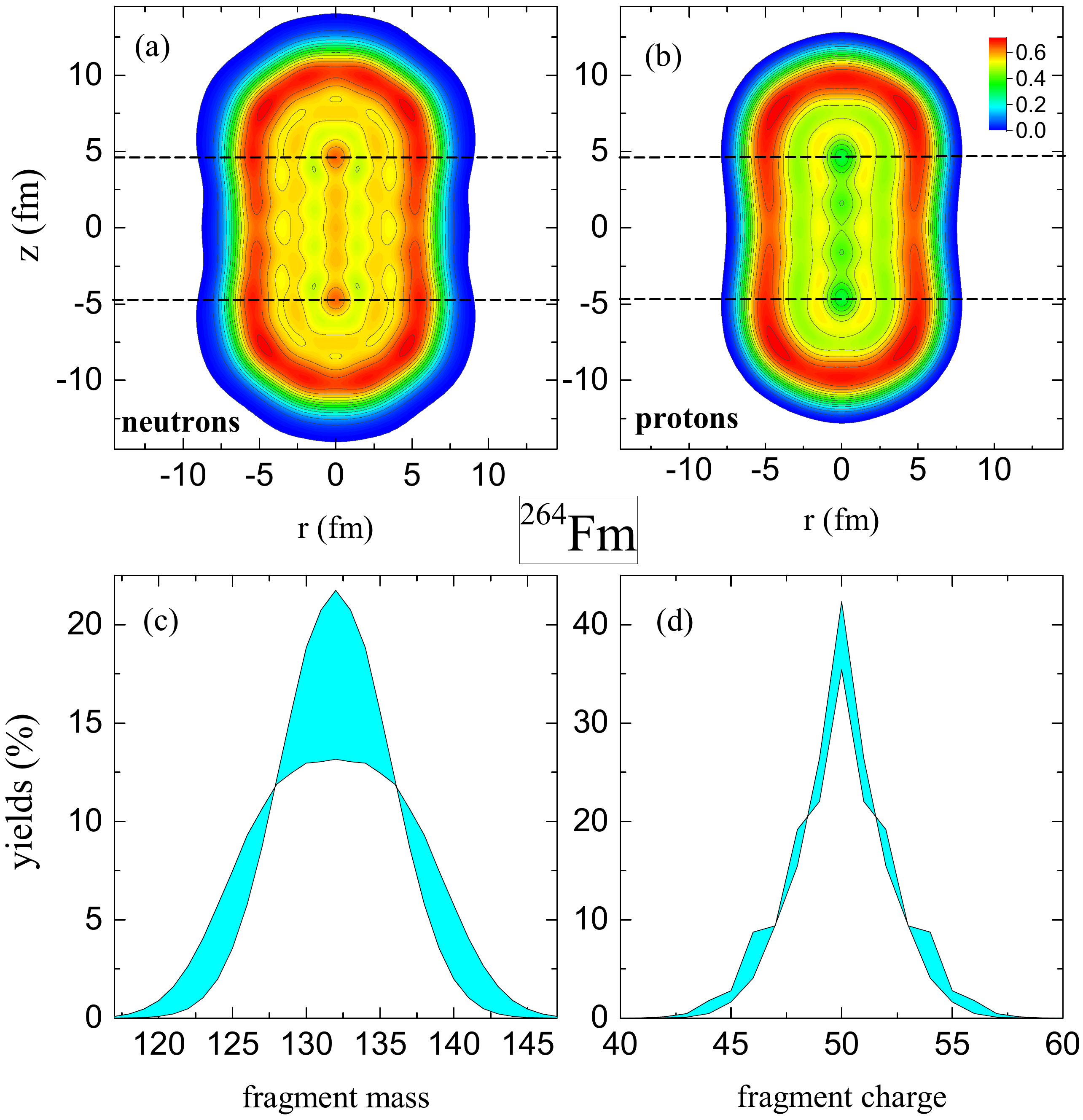}
	\caption{Fission properties of $^{264}$Fm. Nucleon localization functions for neutrons (a) and protons (b) calculated at the pre-scission configuration. Dashed lines mark the prefragment centers. The predicted mass and charge fragment distributions are shown in panels (c) and (d), respectively.}
	\label{fig:fm_plot}
\end{figure}

We should mention here that the microscopic and dynamical effects are accounted for during the process of selecting the pre-scission configuration and in defining the prefragments using NLFs. Our methodology can thus be viewed as a hybrid method where a  microscopic technique is applied in conjunction with a  statistical ansatz in two different domains of the configuration space. This technique is quite robust {\it even} for systems where a prominent neck does not appear. Such a scenario may be at play for a highly fissile system such as $^{264}$Fm, as it is shown by the neutron and proton localization functions shown in Figs.~\hyperref[fig:fm_plot]{\ref{fig:fm_plot}(a)} and \hyperref[fig:fm_plot]{\ref{fig:fm_plot}(b)}, respectively.
Although in this case a neck is not developed at the pre-scission configuration, we find that the prefragments have well-defined centers at $N/Z=49/78$, which allows for a proper identification of the proton and neutron number of fission prefragments \cite{zhang2016}.  Interestingly, the calculated charge distribution for $^{264}$Fm shown in Fig.~\hyperref[fig:fm_plot]{\ref{fig:fm_plot}(d)} is strongly peaked around $Z=50$ and exhibits no OES. This is because the prefragments strongly favor the symmetric fission into two doubly-magic $^{132}$Sn fragments; hence the number of neck nucleons available for redistribution (2 protons and 8 neutrons) is limited.

\begin{figure}[tb]
	\includegraphics[width=\columnwidth]{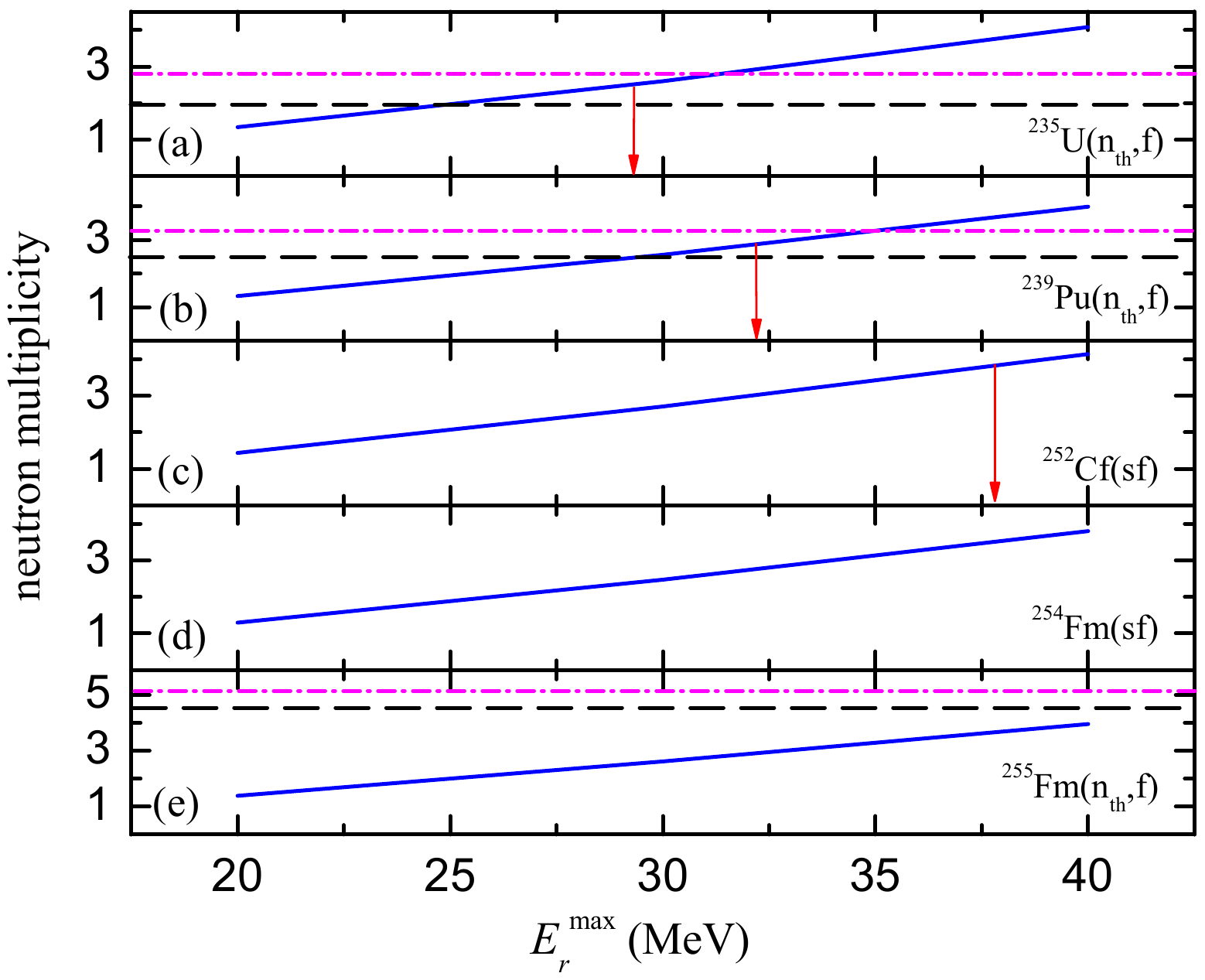}
	\caption{Total neutron multiplicities $\overline{\nu}_{tot}$ for different fissioning systems as a function of $E_r^{\text{max}}$. Dashed and dash-dotted lines are $\overline{\nu}_{tot}$ obtained from the TALYS code~\cite{koning2007} for $E_{n_\textrm{th}}=1$~eV using the Hauser-Feshbach formalism and the GEF fission yields model~\cite{schmidt2016}, respectively. The values of $E_r^{\text{max}}$ corresponding to measured values of $\overline{\nu}_{tot}$~\cite{kor03} are indicated by arrows.}
	\label{fig:nmul}
\end{figure}
In order to asses the robustness of the neutron evaporation scheme adopted here, Fig.~\ref{fig:nmul} shows the calculated total neutron multiplicities ($\overline{\nu}_{tot}$) for different fissioning systems as a function of $E_r^{\text{max}}$. As expected, $\overline{\nu}_{tot}$ increases with $E_r^{\text{max}}$, in agreement with experimental findings. However, as demonstrated in Fig.~\ref{fig:Etest}, yield distributions are not sensitive to $E_r^{\text{max}}$ within the range suggested by experimental $\overline{\nu}_{tot}$. For neutron-induced fission, we compare our predicted $\overline{\nu}_{tot}$ values with the results obtained from two different prescriptions: the TALYS~1.95 code~\cite{koning2007} that employs the traditional Hauser-Feshbach formalism and the GEF fission yields model~\cite{schmidt2016}. As shown in Fig.~\ref{fig:nmul}, we find that our model and TALYS agree within the $E_r^\text{max}$ range suggested by experimental $\overline{\nu}_{tot}$.

\subsection{OES in fission fragment yields}
\begin{figure}[tb]
	\includegraphics[width=\columnwidth]{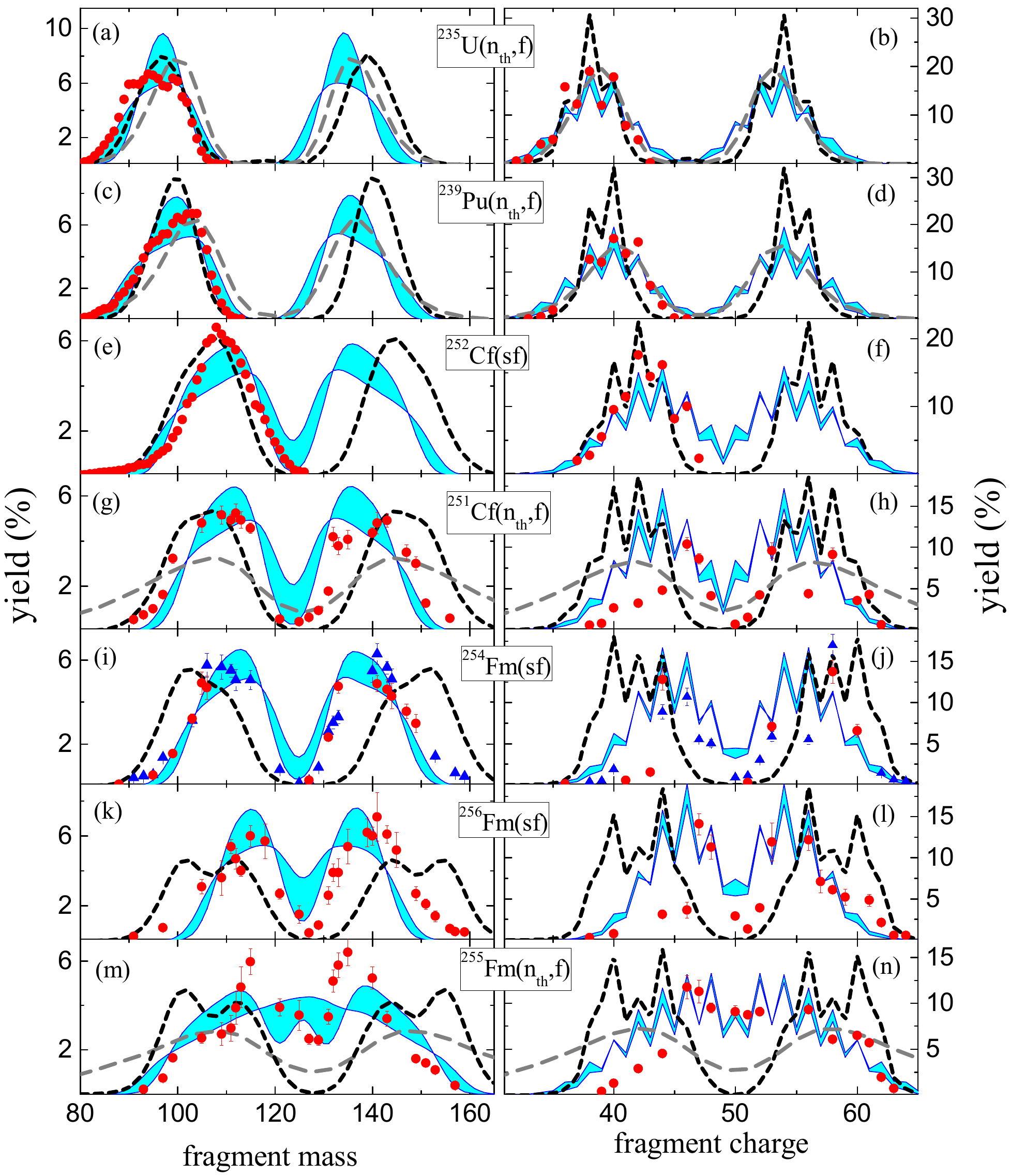}
	\caption{Calculated secondary (post-neutron emission) mass (left panels) and charge (right panels) fission fragment distributions using our model 
		(blue bands), the BSM~\cite{Mumpower2020} (gray dashed lines), and
	the SPM~\cite{lemaitre2021} (black dashed lines) models. Red (circle) and blue (triangle) symbols show experimental data: (a)-(b)~\cite{lang1980}; (c)-(d)~\cite{laidler1962}; (e)~\cite{zeynalov2011}, (f)~\cite{mario1981}; (g)-(h)~\cite{flynn1975a}; (i)-(j)~\cite{harbour1973,gindler1977}; (k)-(l)~\cite{flynn1972}; (m)-(n)~\cite{flynn1975a} (Only light-fragment data are available for (a)-(f)).}
	\label{fig:actinides_ffds}
\end{figure}

After validating our model with respect to possible variations in the model inputs, we now focus on the OES effect incorporated according to Eq.~\eqref{eq:pair}. Figure~\ref{fig:actinides_ffds} shows the mass and charge yields for selected nuclei.
Both the mass and charge yields are measured for these nuclei at low excitation energies (thermal and spontaneous fission) where OES is expected to be most prominent. Modifications due the variations of the pairing term (\ref{eq:pair}) and the effect of the neutron evaporation are discussed in Section~\ref{sec:neut_evap}.  Broadening in the distributions of \figc{fig:actinides_ffds} is associated with the two-particle uncertainty~\cite{Sadhukhan2020}. In general, the agreement of our  mass and charge distributions with experiment is quite satisfactory: the experimental peak locations and distribution widths are reproduced, and so is the OES in charge yields in most of the nuclei considered. We recall that in our model the physical mechanisms responsible for these three observables are very different. Namely, the peak position is mainly affected by microscopic shell effects~\cite{staszczak2009,scamps2018a};  the width of the distribution is driven by stochastic dynamics~\cite{sadhukhan2016}; and the OES in charge distributions can be understood in terms of the statistical formation of fragments with an odd number of protons being hindered by pairing correlations. We notice that somehow larger discrepancies are found for systems that exhibit asymmetries between light and heavy charge distributions, with the light fragment charge being overestimated. This result may suggest the occurrence of beta decay in light fragments that has not been accounted for by our model. However, more detailed and accurate experimental data are needed in order to draw firm conclusions since in some cases, such as $^{254}$Fm, different experimental results are not consistent. 

Left panels of \figc{fig:actinides_ffds} show that the OES is absent in experimental mass distributions. This quenching can be related to two distinct effects. First, averaging over contributions from different isotopes and isobars suppresses OES. We verified this in the mass distributions of primary fragments. The secondary mass distributions are further smoothed out due to neutron evaporation from excited fission fragments (see discussion in Section~\ref{sec:neut_evap}). For completeness, \figc{fig:actinides_ffds} shows comparison with the BSM~\cite{Mumpower2020} and SPM~\cite{lemaitre2021}  predictions. As mentioned above, the BSM method accounts for the dissipative effects required to properly describe widths of the fission fragment distributions. The absence of OES in charge-yield distributions predicted by BSM is not surprising. Indeed, this model lacks the pair-braking mechanism and the charge yields are obtained by simply rescaling the mass yields. While this limitation can be circumvented  by either introducing charge asymmetry as an additional degree of freedom~\cite{moller2015c} or by means of particle number projection~\cite{Verriere2021b},  such extensions have not yet been used in large-scale calculations. 
The SPM calculations reproduce the experimental OES in the charge yields of lighter actinides, but the agreement gets worse for fermium isotopes. 

\begin{figure}[tbh]
	\centering\includegraphics[width=\columnwidth]{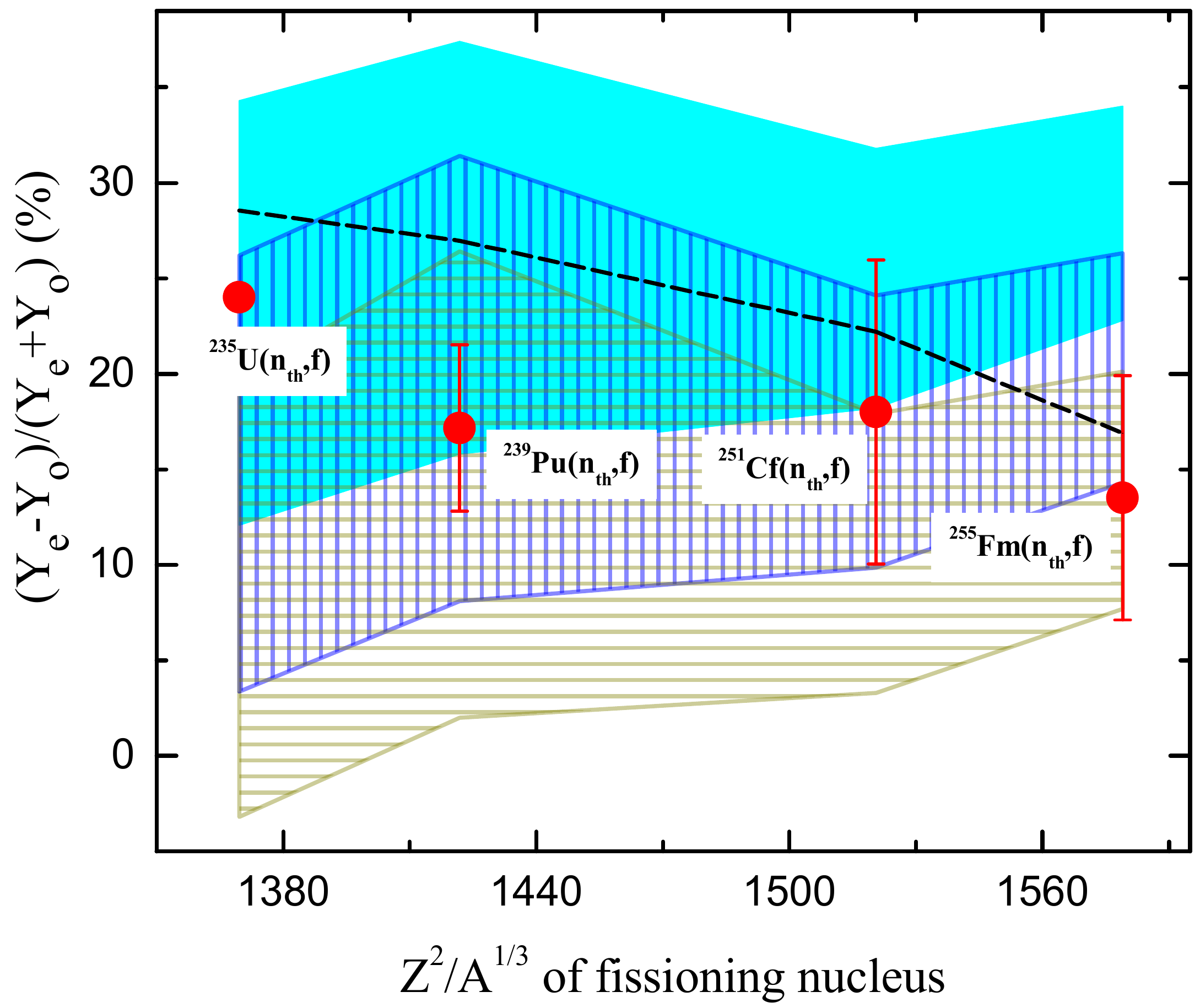}
	\caption{Odd-even difference $\delta_Y$ in charge yields for $\alpha= 0.31$ (blue band), 0.41 (vertical pattern), 0.51 (horizontal pattern) as a function of the Coulomb factor $Z^2/A^{1/3}$. Red dots show experimental data~\cite{lang1980,laidler1962,flynn1975a}. Dashed line is $\delta_Y$ obtained from the SPM~\cite{lemaitre2021}.}
	\label{fig:delp}
\end{figure}
The odd-even differences in charge yield are sensitive to the average pairing energy $\tilde{\Delta}$. It can be characterized in terms of  the odd-even difference $\delta_Y \equiv (Y_e-Y_o)/(Y_e+Y_o)$ \cite{bocquet1989}, where $Y_e$ ($Y_o$) is the total yield of even-$Z$ (odd-$Z$) fragments. The quantity $\delta_Y$ is very sensitive to the paring strength and it is expected to decrease exponentially with the Coulomb parameter $Z^2/A^{1/3}$ of the fissioning system. We calculated $\delta_Y$ for $\alpha=0.31$ in Eq.~\eqref{eq:pair}  and two other values (0.41 and 0.51) as well. The corresponding yields are compared with the experimental data in Fig.~\ref{fig:delp}. Evidently, $\delta_Y$ is broad due to the two-particle uncertainty. Although $\alpha=0.41$ seems to agree better with the experimental $\delta_Y$, a larger data set with wide variations of $Z^2/A^{1/3}$ is needed to fine-tune the average pairing energy. In the present work, we therefore stick to the original value $\alpha=0.31$~\cite{bertsch2009}. We should also mention here that $\delta_Y$ does not uniquely determine the quality of a model. For example, $\delta_Y$ from the SPM are close to the experimental $\delta_Y$ for $^{251}$Cf$(n_\textrm{th},f)$ and $^{255}$Fm$(n_\textrm{th},f)$ even though the corresponding charge-yield distributions are quite distinct (see Fig.~\ref{fig:actinides_ffds}).

\begin{figure}[tb]
\includegraphics[width=\columnwidth]{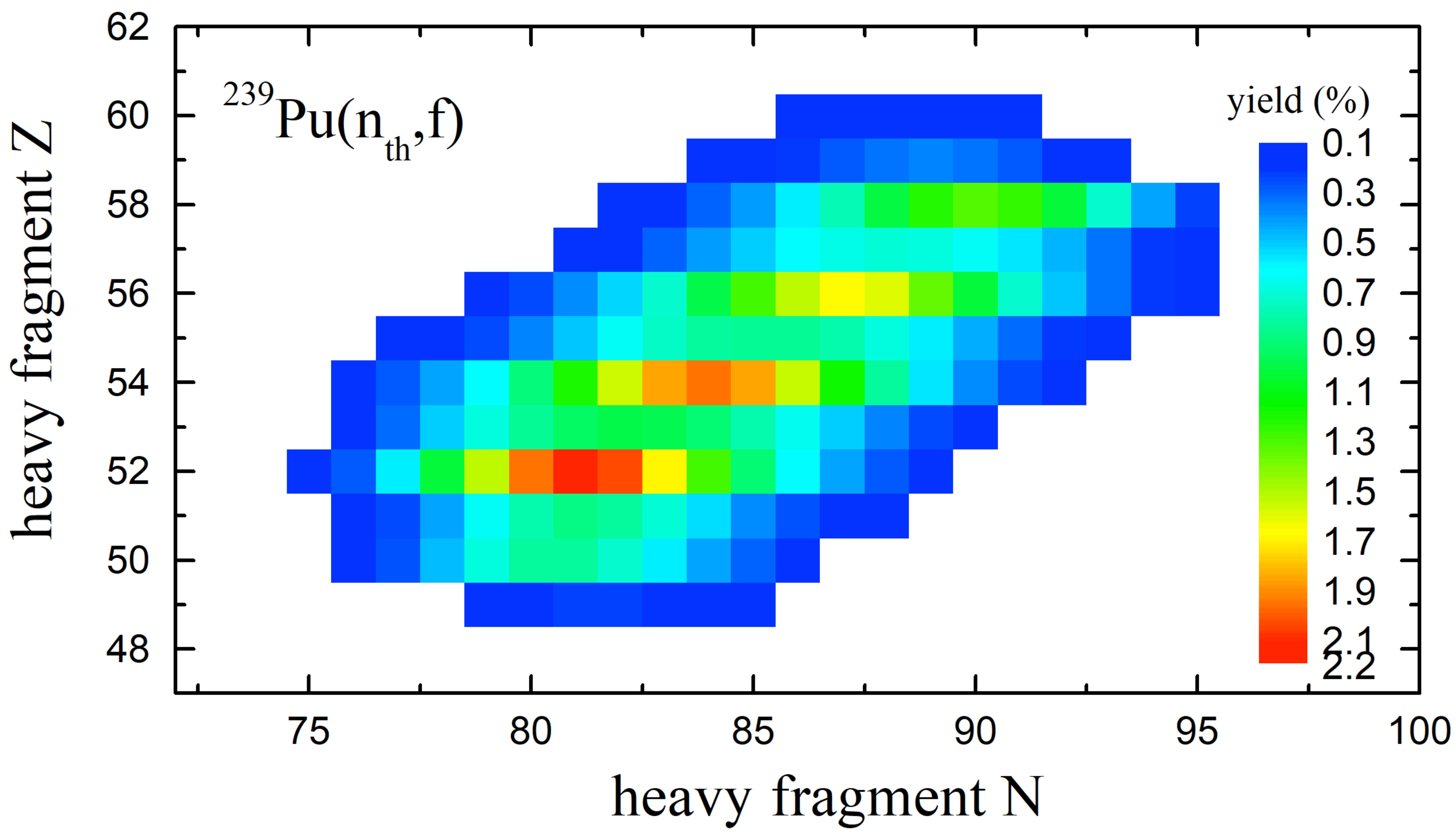}
	\caption{Distribution of heavy fragments in the $N\text{-}Z$ plane calculated in our model.}
	\label{fig:n-z-distribution}
\end{figure}

\begin{figure}[tb]
\includegraphics[width=\columnwidth]{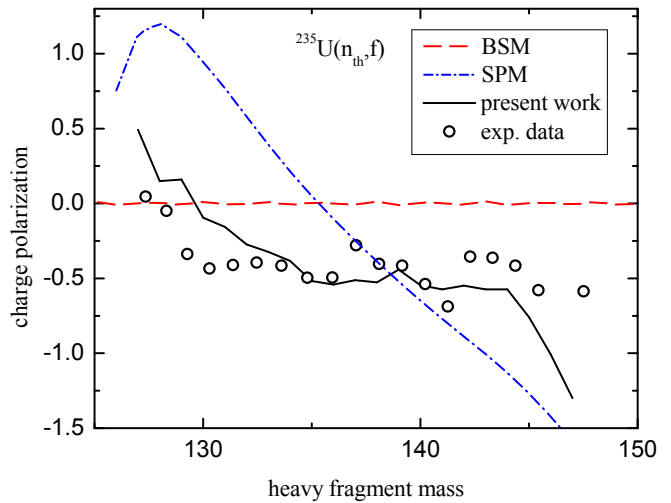}
	\caption{Predicted charge polarization of $^{235}\textrm{U}(n_\textrm{th},f)$ heavy fragment from our model (solid line), BSM model (dashed line), and SPM (dash dotted line). Open circles show experimental data~\cite{lang1980}.}
	\label{fig:ch-pol}
\end{figure}

Besides mass and charge distributions, different characterizations of the fragment yields can be found in the literature. To further assess the precision and accuracy of our model, we present some complementary results for mass and charge yields. Figure~\ref{fig:n-z-distribution} shows the predicted $^{239} \textrm{Pu} (n_\textrm{th},f)$ fragment distribution in the $N\text{-}Z$ plane, where the OES in charge yields is clearly visible. We find that the distribution predicted by our model is broader along the $N$-axis compared to the recent macroscopic-microscopic calculations~\cite{schmitt2021}. 
Another useful quantity is the charge polarization of fragments measured in terms of $\langle Z \rangle - Z_{UCD}$, where $Z_{UCD}$ is the isospin-unchanged charge distribution~\cite{iyer1971}. Calculated charge polarization of $^{235}\textrm{U}(n_\textrm{th},f)$ heavy fragments is compared with the experimental data in Fig.~\ref{fig:ch-pol}. Except for the most asymmetric configurations with very small  yields, we find good overall agreement with experiment. Moreover, our predictions are closer to the measured data compared to the SPM and BSM results. In case of BSM, the magnitude of charge polarization is virtually zero since charge distributions are obtained by rescaling the mass yields, which is equivalent to the expression of charge yields given by the unchanged charge density $Z_{UCD}$.  
%
%
\subsection{Impact of neutron evaporation on OES}\label{sec:neut_evap}
\begin{figure}[tb]
	\includegraphics[width=\columnwidth]{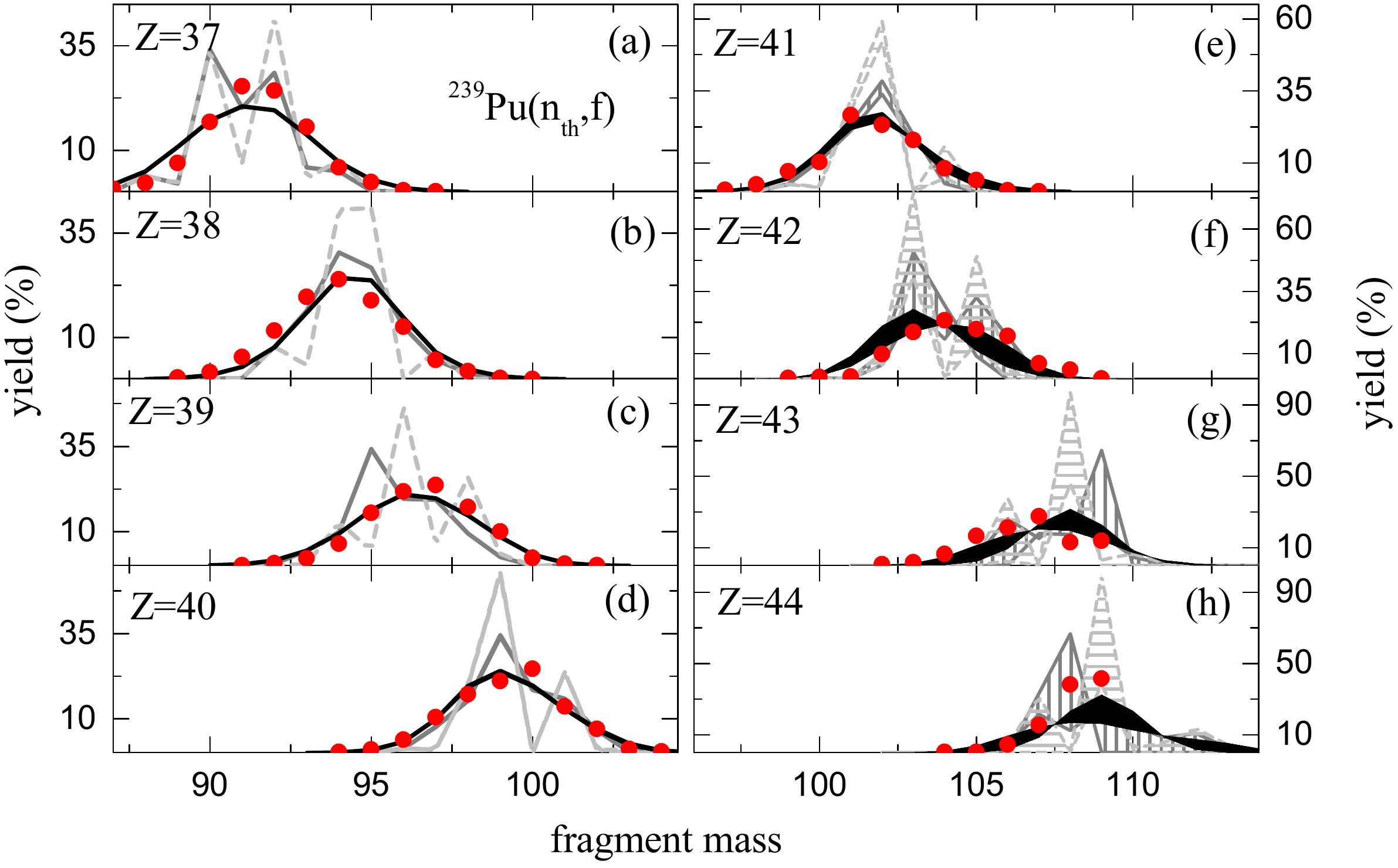}
	\caption{Partial fragment yields  for  fission of $^{240}$Pu induced by thermal neutrons, $^{239}$Pu(n$_{\rm th}$,f), for different fragment isotopes. Secondary (following neutron evaporation) yields calculated for a single $E_r^\text{max}=32$ MeV (light gray lines and horizontal patterns),  average $\langle E_r^\text{max}\rangle=32$ MeV (gray lines and vertical patterns), and subsequent Gaussian convolution with $\sigma=1$ (black lines and bands) are compared with experimental data (symbols)~\cite{schmitt1984}.}
	\label{fig:neut_evap}
\end{figure}

The absence of OES in experimental mass yields of secondary fission fragments can be traced back to neutron evaporation from the excited fission fragments~\cite{schmitt1984} in which nucleonic pairing is quenched because of thermal excitations \cite{Langanke1996,Kaneko2004}. To analyze this effect, we study the impact of neutron evaporation on different isotopes of fission fragments. Figure~\ref{fig:neut_evap} shows the isotopic fragment yields of $^{240}$Pu predicted by our model and compares them  to experiment~\cite{schmitt1984}. Interestingly, OES survives in the secondary isotopic mass distributions when the fluctuations due to dissipative effects are neglected, i.e., when a fixed value of $E_r$ is assumed for a particular mode of fragmentation (here we took $E_r^\text{max}=32$\,MeV reproducing the experimental neutron multiplicity). However, as  explained in Sec.~\ref{sec:formalism},  $E_\textrm{presc}$ and $E_r$ are expected to fluctuate due to the presence of dissipative energy transfer. This is supported by the experimental finding that the TKE per fragment mass shows a $15-30$~MeV variation~\cite{declercq1536,schmitt1984}, suggesting a spread in $E_C+E_\textrm{presc}$ which in turn results in a fluctuation of $E_r$ (see also Eq.~\eqref{eq:enrg} and Fig.~\ref{fig:schematic}). To take this effect into account, we computed the fission fragments assuming a spread in $E_r^{\text{max}}$ of 8 MeV (in accordance with the energy fluctuation caused by fluctuation-dissipation of collective kinetic energy~\cite{Caamano2017}). As illustrated in \figc{fig:neut_evap}, in the average secondary mass yields corresponding to $E_r^\text{max}=$28, 32, and 36~MeV, the OES is reduced considerably, improving  the agreement with experimental data. In general, yield patterns should shift toward lower masses with increasing $E_r^{\text{max}}$ since larger excitation energy facilitates neutron evaporation (see Fig.~\ref{fig:nmul}). However this mechanism strongly depends on the relative difference between $E_r$ and $S_n$, which must be positive in order to allow for neutron emission. 
We shall mention that, even though our simple procedure results in a reasonable description of secondary isotopic fission fragments, more precise calculations may be in order to obtain a better agreement with  experiment. For completeness, we show in \figc{fig:neut_evap} that a Gaussian smoothing of the averaged results with $\sigma=1$  reproduces the experimental yields, thus establishing a more appropriate way to incorporate fluctuations.

\begin{figure}[tb]
	\includegraphics[width=\columnwidth]{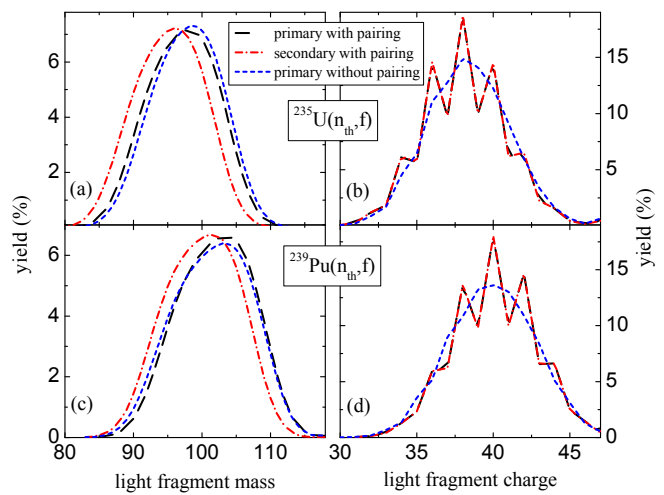}
	\caption{Fission fragment mass (left panels) and charge (right panels) yields calculated for $E_r^{\text{\text{max}}}=40$ MeV. Primary yields are obtained with (short-dashed lines) and without (long-dashed lines) the pairing term, and secondary yields (dash-dotted lines) are obtained with the pairing term.}
	\label{fig:pair-evap}
\end{figure}

Finally, Fig.~\ref{fig:pair-evap} illustrates the interplay between pairing and neutron evaporation. As expected, secondary  fragment-mass yields following neutron evaporation are shifted towards lower masses. Also, no OES is observed in the charge yields if the pairing term \eqref{eq:pair} is absent. 
%
%
\subsection{OES in exotic nuclei}
\begin{figure}[tbh]
	\includegraphics[width=\columnwidth]{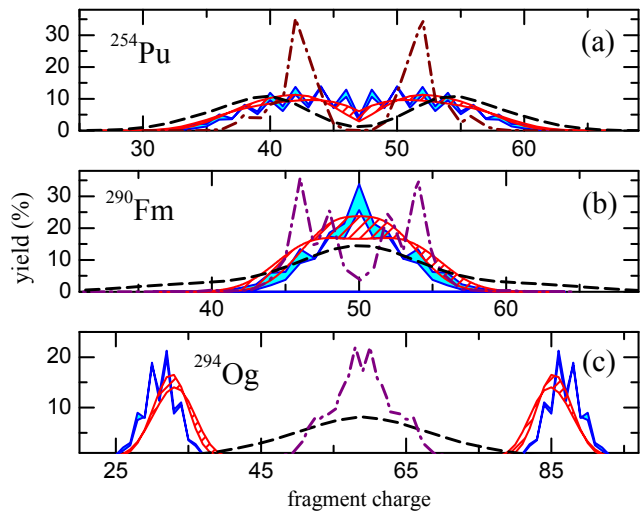}
	\caption{Fragment charge distributions of $^{254}$Pu, $^{290}$Fm, 
	               and $^{294}$Og obtained in this work (blue bands) and predicted in Ref.~\cite{Sadhukhan2020} by neglecting 
	               OES (red dashed bands). Predictions of BSM~\cite{Mumpower2020} and SPM~\cite{lemaitre2021} models 
	               are shown by dashed and dash-dotted lines, respectively.}
	\label{fig:exotic_ffds}
\end{figure}

We conclude this study by computing the fragment charge distributions of three exotic systems: the \rpa\ nuclei $^{254}_{\hphantom{0}94}$Pu and $^{290}_{100}$Fm~\cite{vassh2019,giuliani2019a}, and the superheavy system $^{294}_{118}$Og. Figure~\ref{fig:exotic_ffds} shows the fragment charge distributions predicted for these three nuclei compared with our earlier  results where OES effects have been neglected~\cite{Sadhukhan2020}, and  with the results of  BSM~\cite{Mumpower2020} and SPM~\cite{lemaitre2021} models. 
While both our model and BSM  predict a broad asymmetric distribution for $^{254}$Pu, SPM yields a rather narrow distribution with sharp maxima. We notice that the emergence of OES in our model has some impact on the charge distributions of $^{290}$Fm and $^{294}$Og. Namely, in the case of  $^{290}$Fm, the probability of a symmetric split into two tin isotopes is increased by  $\sim$10\%. This results in a narrowing of the width, thus increasing the discrepancy between our prediction and the broad distribution obtained with the BSM model. Interestingly, SPM predicts asymmetric distribution for $^{290}$Fm. 

For $^{294}$Og, microscopic models~\cite{warda2018,matheson2019}  predict highly asymmetric fission, or  cluster emission, with a heavy fragment near the doubly magic $^{208}$Pb. Such a mode is clearly seen in our calculations. Again, the appearance of the  OES results in a shift towards more asymmetric configurations  resulting in  a charge distribution centered around Rn isotopes for the heavy fragments. The BSM and SPM predictions are strikingly different. Clearly,  experimental search for a cluster emission from  $^{294}$Og would be of great value.

\section{Conclusions}
We have developed a  microscopic-statistical model of charge and mass fission-yield distributions.  The proposed hybrid approach contains no parameters adjusted to  fission-fragment data for individual nuclei.   Our predictions of OES in charge yields explain measured values for a wide range of fissioning nuclei, as well as  experimental data on widths and peak positions of the fission fragment distributions. This encouraging result  supports our assumption that that  microscopic rearrangements of nucleons into prefragments  occur well before scission, and the subsequent dynamics is mainly driven by the thermal excitations and bulk features  of the nuclear binding.

We explored the impact of neutron evaporation on quenching of the OES  observed in mass distributions, and found that experimental results can be reproduced assuming a simple emission scheme. Finally, we studied the fission fragment distributions of exotic nuclei, and show that the OES can impact the charge yields of such systems. The application of our model to large-scale simulations of \rpa\ nucleosynthesis is in progress.

\begin{acknowledgments} 
The authors are grateful to J.-F.~Lema{\^i}tre and S.~Goriely for providing the updated SPY2 results showed in Figs.~\ref{fig:actinides_ffds} and~\ref{fig:exotic_ffds}, and for elucidations about the SPM model. This work was  supported by the U.S. Department of Energy under Award Numbers DOE-DE-NA0003885  (NNSA, the Stewardship Science Academic Alliances program), DE-SC0013365 (Office of Science), and DE-SC0018083 (Office of Science, NUCLEI SciDAC-4 collaboration) and the Caritro Foundation (Rif. Int.: 2020.0259). 
\end{acknowledgments}

\bibliography{main}

\begin{thebibliography}{74}%
\makeatletter
\providecommand \@ifxundefined [1]{%
 \@ifx{#1\undefined}
}%
\providecommand \@ifnum [1]{%
 \ifnum #1\expandafter \@firstoftwo
 \else \expandafter \@secondoftwo
 \fi
}%
\providecommand \@ifx [1]{%
 \ifx #1\expandafter \@firstoftwo
 \else \expandafter \@secondoftwo
 \fi
}%
\providecommand \natexlab [1]{#1}%
\providecommand \enquote  [1]{``#1''}%
\providecommand \bibnamefont  [1]{#1}%
\providecommand \bibfnamefont [1]{#1}%
\providecommand \citenamefont [1]{#1}%
\providecommand \href@noop [0]{\@secondoftwo}%
\providecommand \href [0]{\begingroup \@sanitize@url \@href}%
\providecommand \@href[1]{\@@startlink{#1}\@@href}%
\providecommand \@@href[1]{\endgroup#1\@@endlink}%
\providecommand \@sanitize@url [0]{\catcode `\\12\catcode `\$12\catcode
  `\&12\catcode `\#12\catcode `\^12\catcode `\_12\catcode `\%12\relax}%
\providecommand \@@startlink[1]{}%
\providecommand \@@endlink[0]{}%
\providecommand \url  [0]{\begingroup\@sanitize@url \@url }%
\providecommand \@url [1]{\endgroup\@href {#1}{\urlprefix }}%
\providecommand \urlprefix  [0]{URL }%
\providecommand \Eprint [0]{\href }%
\providecommand \doibase [0]{http://dx.doi.org/}%
\providecommand \selectlanguage [0]{\@gobble}%
\providecommand \bibinfo  [0]{\@secondoftwo}%
\providecommand \bibfield  [0]{\@secondoftwo}%
\providecommand \translation [1]{[#1]}%
\providecommand \BibitemOpen [0]{}%
\providecommand \bibitemStop [0]{}%
\providecommand \bibitemNoStop [0]{.\EOS\space}%
\providecommand \EOS [0]{\spacefactor3000\relax}%
\providecommand \BibitemShut  [1]{\csname bibitem#1\endcsname}%
\let\auto@bib@innerbib\@empty
\bibitem [{\citenamefont {Vogel}\ \emph {et~al.}(2015)\citenamefont {Vogel},
  \citenamefont {Wen},\ and\ \citenamefont {Zhang}}]{vogel2015}%
  \BibitemOpen
  \bibfield  {author} {\bibinfo {author} {\bibfnamefont {P.}~\bibnamefont
  {Vogel}}, \bibinfo {author} {\bibfnamefont {L.}~\bibnamefont {Wen}}, \ and\
  \bibinfo {author} {\bibfnamefont {C.}~\bibnamefont {Zhang}},\ }\bibfield
  {title} {\enquote {\bibinfo {title} {{Neutrino oscillation studies with
  reactors}},}\ }\href {\doibase 10.1038/ncomms7935} {\bibfield  {journal}
  {\bibinfo  {journal} {Nat. Commun.}\ }\textbf {\bibinfo {volume} {6}},\
  \bibinfo {pages} {6935} (\bibinfo {year} {2015})}\BibitemShut {NoStop}%
\bibitem [{\citenamefont {Horowitz}\ \emph {et~al.}(2019)\citenamefont
  {Horowitz}, \citenamefont {Arcones}, \citenamefont {C{\^{o}}t{\'{e}}},
  \citenamefont {Dillmann}, \citenamefont {Nazarewicz}, \citenamefont
  {Roederer}, \citenamefont {Schatz}, \citenamefont {Aprahamian}, \citenamefont
  {Atanasov}, \citenamefont {Bauswein}, \citenamefont {Beers}, \citenamefont
  {Bliss}, \citenamefont {Brodeur}, \citenamefont {Clark}, \citenamefont
  {Frebel}, \citenamefont {Foucart}, \citenamefont {Hansen}, \citenamefont
  {Just}, \citenamefont {Kankainen}, \citenamefont {McLaughlin}, \citenamefont
  {Kelly}, \citenamefont {Liddick}, \citenamefont {Lee}, \citenamefont
  {Lippuner}, \citenamefont {Martin}, \citenamefont {Mendoza-Temis},
  \citenamefont {Metzger}, \citenamefont {Mumpower}, \citenamefont
  {Perdikakis}, \citenamefont {Pereira}, \citenamefont {O'Shea}, \citenamefont
  {Reifarth}, \citenamefont {Rogers}, \citenamefont {Siegel}, \citenamefont
  {Spyrou}, \citenamefont {Surman}, \citenamefont {Tang}, \citenamefont
  {Uesaka},\ and\ \citenamefont {Wang}}]{horowitz2018}%
  \BibitemOpen
  \bibfield  {author} {\bibinfo {author} {\bibfnamefont {C.~J.}\ \bibnamefont
  {Horowitz}}, \bibinfo {author} {\bibfnamefont {A.}~\bibnamefont {Arcones}},
  \bibinfo {author} {\bibfnamefont {B.}~\bibnamefont {C{\^{o}}t{\'{e}}}},
  \bibinfo {author} {\bibfnamefont {I.}~\bibnamefont {Dillmann}}, \bibinfo
  {author} {\bibfnamefont {W.}~\bibnamefont {Nazarewicz}}, \bibinfo {author}
  {\bibfnamefont {I.~U.}\ \bibnamefont {Roederer}}, \bibinfo {author}
  {\bibfnamefont {H.}~\bibnamefont {Schatz}}, \bibinfo {author} {\bibfnamefont
  {A.}~\bibnamefont {Aprahamian}}, \bibinfo {author} {\bibfnamefont
  {D.}~\bibnamefont {Atanasov}}, \bibinfo {author} {\bibfnamefont
  {A.}~\bibnamefont {Bauswein}}, \bibinfo {author} {\bibfnamefont {T.~C.}\
  \bibnamefont {Beers}}, \bibinfo {author} {\bibfnamefont {J.}~\bibnamefont
  {Bliss}}, \bibinfo {author} {\bibfnamefont {M.}~\bibnamefont {Brodeur}},
  \bibinfo {author} {\bibfnamefont {J.~A.}\ \bibnamefont {Clark}}, \bibinfo
  {author} {\bibfnamefont {A.}~\bibnamefont {Frebel}}, \bibinfo {author}
  {\bibfnamefont {F.}~\bibnamefont {Foucart}}, \bibinfo {author} {\bibfnamefont
  {C.~J.}\ \bibnamefont {Hansen}}, \bibinfo {author} {\bibfnamefont
  {O.}~\bibnamefont {Just}}, \bibinfo {author} {\bibfnamefont {A.}~\bibnamefont
  {Kankainen}}, \bibinfo {author} {\bibfnamefont {G.~C.}\ \bibnamefont
  {McLaughlin}}, \bibinfo {author} {\bibfnamefont {J.~M.}\ \bibnamefont
  {Kelly}}, \bibinfo {author} {\bibfnamefont {S.~N.}\ \bibnamefont {Liddick}},
  \bibinfo {author} {\bibfnamefont {D.~M.}\ \bibnamefont {Lee}}, \bibinfo
  {author} {\bibfnamefont {J.}~\bibnamefont {Lippuner}}, \bibinfo {author}
  {\bibfnamefont {D.}~\bibnamefont {Martin}}, \bibinfo {author} {\bibfnamefont
  {J.}~\bibnamefont {Mendoza-Temis}}, \bibinfo {author} {\bibfnamefont {B.~D.}\
  \bibnamefont {Metzger}}, \bibinfo {author} {\bibfnamefont {M.~R.}\
  \bibnamefont {Mumpower}}, \bibinfo {author} {\bibfnamefont {G.}~\bibnamefont
  {Perdikakis}}, \bibinfo {author} {\bibfnamefont {J.}~\bibnamefont {Pereira}},
  \bibinfo {author} {\bibfnamefont {B.~W.}\ \bibnamefont {O'Shea}}, \bibinfo
  {author} {\bibfnamefont {R.}~\bibnamefont {Reifarth}}, \bibinfo {author}
  {\bibfnamefont {A.~M.}\ \bibnamefont {Rogers}}, \bibinfo {author}
  {\bibfnamefont {D.~M.}\ \bibnamefont {Siegel}}, \bibinfo {author}
  {\bibfnamefont {A.}~\bibnamefont {Spyrou}}, \bibinfo {author} {\bibfnamefont
  {R.}~\bibnamefont {Surman}}, \bibinfo {author} {\bibfnamefont
  {X.}~\bibnamefont {Tang}}, \bibinfo {author} {\bibfnamefont {T.}~\bibnamefont
  {Uesaka}}, \ and\ \bibinfo {author} {\bibfnamefont {M.}~\bibnamefont
  {Wang}},\ }\bibfield  {title} {\enquote {\bibinfo {title} {R-process
  nucleosynthesis: connecting rare-isotope beam facilities with the cosmos},}\
  }\href {\doibase 10.1088/1361-6471/ab0849} {\bibfield  {journal} {\bibinfo
  {journal} {J. Phys. G}\ }\textbf {\bibinfo {volume} {46}},\ \bibinfo {pages}
  {083001} (\bibinfo {year} {2019})}\BibitemShut {NoStop}%
\bibitem [{\citenamefont {Cowan}\ \emph {et~al.}(2021)\citenamefont {Cowan},
  \citenamefont {Sneden}, \citenamefont {Lawler}, \citenamefont {Aprahamian},
  \citenamefont {Wiescher}, \citenamefont {Langanke}, \citenamefont
  {Mart{\'{i}}nez-Pinedo},\ and\ \citenamefont {Thielemann}}]{Cowan2021}%
  \BibitemOpen
  \bibfield  {author} {\bibinfo {author} {\bibfnamefont {J.~J.}\ \bibnamefont
  {Cowan}}, \bibinfo {author} {\bibfnamefont {C.}~\bibnamefont {Sneden}},
  \bibinfo {author} {\bibfnamefont {J.~E.}\ \bibnamefont {Lawler}}, \bibinfo
  {author} {\bibfnamefont {A.}~\bibnamefont {Aprahamian}}, \bibinfo {author}
  {\bibfnamefont {M.}~\bibnamefont {Wiescher}}, \bibinfo {author}
  {\bibfnamefont {K.}~\bibnamefont {Langanke}}, \bibinfo {author}
  {\bibfnamefont {G.}~\bibnamefont {Mart{\'{i}}nez-Pinedo}}, \ and\ \bibinfo
  {author} {\bibfnamefont {F.-K.}\ \bibnamefont {Thielemann}},\ }\bibfield
  {title} {\enquote {\bibinfo {title} {{Origin of the heaviest elements: The
  rapid neutron-capture process}},}\ }\href {\doibase
  10.1103/RevModPhys.93.015002} {\bibfield  {journal} {\bibinfo  {journal}
  {Rev. Mod. Phys.}\ }\textbf {\bibinfo {volume} {93}},\ \bibinfo {pages}
  {015002} (\bibinfo {year} {2021})}\BibitemShut {NoStop}%
\bibitem [{\citenamefont {Bender}\ \emph {et~al.}(2020)\citenamefont {Bender},
  \citenamefont {Bernard}, \citenamefont {Bertsch}, \citenamefont {Chiba},
  \citenamefont {Dobaczewski}, \citenamefont {Dubray}, \citenamefont
  {Giuliani}, \citenamefont {Hagino}, \citenamefont {Lacroix}, \citenamefont
  {Li}, \citenamefont {Magierski}, \citenamefont {Maruhn}, \citenamefont
  {Nazarewicz}, \citenamefont {Pei}, \citenamefont {P{\'{e}}ru}, \citenamefont
  {Pillet}, \citenamefont {Randrup}, \citenamefont {Regnier}, \citenamefont
  {Reinhard}, \citenamefont {Robledo}, \citenamefont {Ryssens}, \citenamefont
  {Sadhukhan}, \citenamefont {Scamps}, \citenamefont {Schunck}, \citenamefont
  {Simenel}, \citenamefont {Skalski}, \citenamefont {Stetcu}, \citenamefont
  {Stevenson}, \citenamefont {Umar}, \citenamefont {Verriere}, \citenamefont
  {Vretenar}, \citenamefont {Warda},\ and\ \citenamefont
  {{\AA}berg}}]{bender2020}%
  \BibitemOpen
  \bibfield  {author} {\bibinfo {author} {\bibfnamefont {M.}~\bibnamefont
  {Bender}}, \bibinfo {author} {\bibfnamefont {R.}~\bibnamefont {Bernard}},
  \bibinfo {author} {\bibfnamefont {G.}~\bibnamefont {Bertsch}}, \bibinfo
  {author} {\bibfnamefont {S.}~\bibnamefont {Chiba}}, \bibinfo {author}
  {\bibfnamefont {J.}~\bibnamefont {Dobaczewski}}, \bibinfo {author}
  {\bibfnamefont {N.}~\bibnamefont {Dubray}}, \bibinfo {author} {\bibfnamefont
  {S.~A.}\ \bibnamefont {Giuliani}}, \bibinfo {author} {\bibfnamefont
  {K.}~\bibnamefont {Hagino}}, \bibinfo {author} {\bibfnamefont
  {D.}~\bibnamefont {Lacroix}}, \bibinfo {author} {\bibfnamefont
  {Z.}~\bibnamefont {Li}}, \bibinfo {author} {\bibfnamefont {P.}~\bibnamefont
  {Magierski}}, \bibinfo {author} {\bibfnamefont {J.}~\bibnamefont {Maruhn}},
  \bibinfo {author} {\bibfnamefont {W.}~\bibnamefont {Nazarewicz}}, \bibinfo
  {author} {\bibfnamefont {J.}~\bibnamefont {Pei}}, \bibinfo {author}
  {\bibfnamefont {S.}~\bibnamefont {P{\'{e}}ru}}, \bibinfo {author}
  {\bibfnamefont {N.}~\bibnamefont {Pillet}}, \bibinfo {author} {\bibfnamefont
  {J.}~\bibnamefont {Randrup}}, \bibinfo {author} {\bibfnamefont
  {D.}~\bibnamefont {Regnier}}, \bibinfo {author} {\bibfnamefont {P.-G.}\
  \bibnamefont {Reinhard}}, \bibinfo {author} {\bibfnamefont {L.~M.}\
  \bibnamefont {Robledo}}, \bibinfo {author} {\bibfnamefont {W.}~\bibnamefont
  {Ryssens}}, \bibinfo {author} {\bibfnamefont {J.}~\bibnamefont {Sadhukhan}},
  \bibinfo {author} {\bibfnamefont {G.}~\bibnamefont {Scamps}}, \bibinfo
  {author} {\bibfnamefont {N.}~\bibnamefont {Schunck}}, \bibinfo {author}
  {\bibfnamefont {C.}~\bibnamefont {Simenel}}, \bibinfo {author} {\bibfnamefont
  {J.}~\bibnamefont {Skalski}}, \bibinfo {author} {\bibfnamefont
  {I.}~\bibnamefont {Stetcu}}, \bibinfo {author} {\bibfnamefont
  {P.}~\bibnamefont {Stevenson}}, \bibinfo {author} {\bibfnamefont
  {S.}~\bibnamefont {Umar}}, \bibinfo {author} {\bibfnamefont {M.}~\bibnamefont
  {Verriere}}, \bibinfo {author} {\bibfnamefont {D.}~\bibnamefont {Vretenar}},
  \bibinfo {author} {\bibfnamefont {M.}~\bibnamefont {Warda}}, \ and\ \bibinfo
  {author} {\bibfnamefont {S.}~\bibnamefont {{\AA}berg}},\ }\bibfield  {title}
  {\enquote {\bibinfo {title} {{Future of nuclear fission theory}},}\ }\href
  {\doibase 10.1088/1361-6471/abab4f} {\bibfield  {journal} {\bibinfo
  {journal} {J. Phys. G}\ }\textbf {\bibinfo {volume} {47}},\ \bibinfo {pages}
  {113002} (\bibinfo {year} {2020})}\BibitemShut {NoStop}%
\bibitem [{\citenamefont {Amiel}\ and\ \citenamefont
  {Feldstein}(1975)}]{amiel1975}%
  \BibitemOpen
  \bibfield  {author} {\bibinfo {author} {\bibfnamefont {S.}~\bibnamefont
  {Amiel}}\ and\ \bibinfo {author} {\bibfnamefont {H.}~\bibnamefont
  {Feldstein}},\ }\bibfield  {title} {\enquote {\bibinfo {title} {Odd-even
  systematics in neutron fission yields of $^{233}\mathrm{U}$ and
  $^{235}\mathrm{U}$},}\ }\href {\doibase 10.1103/PhysRevC.11.845} {\bibfield
  {journal} {\bibinfo  {journal} {Phys. Rev. C}\ }\textbf {\bibinfo {volume}
  {11}},\ \bibinfo {pages} {845} (\bibinfo {year} {1975})}\BibitemShut
  {NoStop}%
\bibitem [{\citenamefont {Lang}\ \emph {et~al.}(1980)\citenamefont {Lang},
  \citenamefont {Clerc}, \citenamefont {Wohlfarth}, \citenamefont {Schrader},\
  and\ \citenamefont {Schmidt}}]{lang1980}%
  \BibitemOpen
  \bibfield  {author} {\bibinfo {author} {\bibfnamefont {W.}~\bibnamefont
  {Lang}}, \bibinfo {author} {\bibfnamefont {H.-G.}\ \bibnamefont {Clerc}},
  \bibinfo {author} {\bibfnamefont {H.}~\bibnamefont {Wohlfarth}}, \bibinfo
  {author} {\bibfnamefont {H.}~\bibnamefont {Schrader}}, \ and\ \bibinfo
  {author} {\bibfnamefont {K.-H.}\ \bibnamefont {Schmidt}},\ }\bibfield
  {title} {\enquote {\bibinfo {title} {Nuclear charge and mass yields for
  $^{235}${U}(n$_\textrm{th}$, f) as a function of the kinetic energy of the
  fission products},}\ }\href {\doibase 10.1016/0375-9474(80)90411-X}
  {\bibfield  {journal} {\bibinfo  {journal} {Nucl. Phys. A}\ }\textbf
  {\bibinfo {volume} {345}},\ \bibinfo {pages} {34} (\bibinfo {year}
  {1980})}\BibitemShut {NoStop}%
\bibitem [{\citenamefont {Mariolopoulos}\ \emph
  {et~al.}(1981{\natexlab{a}})\citenamefont {Mariolopoulos}, \citenamefont
  {Hamelin}, \citenamefont {Blachot}, \citenamefont {Bocquet}, \citenamefont
  {Brissot}, \citenamefont {Crançon}, \citenamefont {Nifenecker},\ and\
  \citenamefont {Ristori}}]{mariolopoulos1981}%
  \BibitemOpen
  \bibfield  {author} {\bibinfo {author} {\bibfnamefont {G.}~\bibnamefont
  {Mariolopoulos}}, \bibinfo {author} {\bibfnamefont {C.}~\bibnamefont
  {Hamelin}}, \bibinfo {author} {\bibfnamefont {J.}~\bibnamefont {Blachot}},
  \bibinfo {author} {\bibfnamefont {J.}~\bibnamefont {Bocquet}}, \bibinfo
  {author} {\bibfnamefont {R.}~\bibnamefont {Brissot}}, \bibinfo {author}
  {\bibfnamefont {J.}~\bibnamefont {Crançon}}, \bibinfo {author}
  {\bibfnamefont {H.}~\bibnamefont {Nifenecker}}, \ and\ \bibinfo {author}
  {\bibfnamefont {C.}~\bibnamefont {Ristori}},\ }\bibfield  {title} {\enquote
  {\bibinfo {title} {{Charge distributions in low-energy nuclear fission and
  their relevance to fission dynamics}},}\ }\href {\doibase
  10.1016/0375-9474(81)90477-2} {\bibfield  {journal} {\bibinfo  {journal}
  {Nucl. Phys. A}\ }\textbf {\bibinfo {volume} {361}},\ \bibinfo {pages} {213}
  (\bibinfo {year} {1981}{\natexlab{a}})}\BibitemShut {NoStop}%
\bibitem [{\citenamefont {Caama{\~{n}}o}\ \emph {et~al.}(2011)\citenamefont
  {Caama{\~{n}}o}, \citenamefont {Rejmund},\ and\ \citenamefont
  {Schmidt}}]{caamano2011}%
  \BibitemOpen
  \bibfield  {author} {\bibinfo {author} {\bibfnamefont {M.}~\bibnamefont
  {Caama{\~{n}}o}}, \bibinfo {author} {\bibfnamefont {F.}~\bibnamefont
  {Rejmund}}, \ and\ \bibinfo {author} {\bibfnamefont {K.-H.}\ \bibnamefont
  {Schmidt}},\ }\bibfield  {title} {\enquote {\bibinfo {title} {{Evidence for
  the predominant influence of the asymmetry degree of freedom on the
  even–odd structure in fission-fragment yields}},}\ }\href {\doibase
  10.1088/0954-3899/38/3/035101} {\bibfield  {journal} {\bibinfo  {journal} {J.
  Phys. G}\ }\textbf {\bibinfo {volume} {38}},\ \bibinfo {pages} {035101}
  (\bibinfo {year} {2011})}\BibitemShut {NoStop}%
\bibitem [{\citenamefont {Mirea}(2014)}]{mirea2014}%
  \BibitemOpen
  \bibfield  {author} {\bibinfo {author} {\bibfnamefont {M.}~\bibnamefont
  {Mirea}},\ }\bibfield  {title} {\enquote {\bibinfo {title} {{Microscopic
  description of the odd-even effect in cold fission}},}\ }\href {\doibase
  10.1103/PhysRevC.89.034623} {\bibfield  {journal} {\bibinfo  {journal} {Phy.
  Rev. C}\ }\textbf {\bibinfo {volume} {89}},\ \bibinfo {pages} {034623}
  (\bibinfo {year} {2014})}\BibitemShut {NoStop}%
\bibitem [{\citenamefont {Mirea}(2017)}]{Mirea2017}%
  \BibitemOpen
  \bibfield  {author} {\bibinfo {author} {\bibfnamefont {M.}~\bibnamefont
  {Mirea}},\ }\bibfield  {title} {\enquote {\bibinfo {title} {Odd-even effect
  dependence on the excitation energy in low energy fission},}\ }\href
  {\doibase 10.1051/epjconf/201714604015} {\bibfield  {journal} {\bibinfo
  {journal} {EPJ Web Conf.}\ }\textbf {\bibinfo {volume} {146}},\ \bibinfo
  {pages} {04015} (\bibinfo {year} {2017})}\BibitemShut {NoStop}%
\bibitem [{\citenamefont {Sadhukhan}\ \emph {et~al.}(2017)\citenamefont
  {Sadhukhan}, \citenamefont {Zhang}, \citenamefont {Nazarewicz},\ and\
  \citenamefont {Schunck}}]{sadhukhan2017}%
  \BibitemOpen
  \bibfield  {author} {\bibinfo {author} {\bibfnamefont {J.}~\bibnamefont
  {Sadhukhan}}, \bibinfo {author} {\bibfnamefont {C.}~\bibnamefont {Zhang}},
  \bibinfo {author} {\bibfnamefont {W.}~\bibnamefont {Nazarewicz}}, \ and\
  \bibinfo {author} {\bibfnamefont {N.}~\bibnamefont {Schunck}},\ }\bibfield
  {title} {\enquote {\bibinfo {title} {Formation and distribution of fragments
  in the spontaneous fission of $^{240}${Pu}},}\ }\href {\doibase
  10.1103/PhysRevC.96.061301} {\bibfield  {journal} {\bibinfo  {journal} {Phys.
  Rev. C}\ }\textbf {\bibinfo {volume} {96}},\ \bibinfo {pages} {061301}
  (\bibinfo {year} {2017})}\BibitemShut {NoStop}%
\bibitem [{\citenamefont {Sierk}(2017)}]{sierk2017}%
  \BibitemOpen
  \bibfield  {author} {\bibinfo {author} {\bibfnamefont {A.~J.}\ \bibnamefont
  {Sierk}},\ }\bibfield  {title} {\enquote {\bibinfo {title} {Langevin model of
  low-energy fission},}\ }\href {\doibase 10.1103/PhysRevC.96.034603}
  {\bibfield  {journal} {\bibinfo  {journal} {Phys. Rev. C}\ }\textbf {\bibinfo
  {volume} {96}},\ \bibinfo {pages} {034603} (\bibinfo {year}
  {2017})}\BibitemShut {NoStop}%
\bibitem [{\citenamefont {Lema{\^{i}}tre}\ \emph {et~al.}(2019)\citenamefont
  {Lema{\^{i}}tre}, \citenamefont {Goriely}, \citenamefont {Hilaire},\ and\
  \citenamefont {Sida}}]{lemaitre2019}%
  \BibitemOpen
  \bibfield  {author} {\bibinfo {author} {\bibfnamefont {J.-F.}\ \bibnamefont
  {Lema{\^{i}}tre}}, \bibinfo {author} {\bibfnamefont {S.}~\bibnamefont
  {Goriely}}, \bibinfo {author} {\bibfnamefont {S.}~\bibnamefont {Hilaire}}, \
  and\ \bibinfo {author} {\bibfnamefont {J.-L.}\ \bibnamefont {Sida}},\
  }\bibfield  {title} {\enquote {\bibinfo {title} {{Fully microscopic
  scission-point model to predict fission fragment observables}},}\ }\href
  {\doibase 10.1103/PhysRevC.99.034612} {\bibfield  {journal} {\bibinfo
  {journal} {Phys. Rev. C}\ }\textbf {\bibinfo {volume} {99}},\ \bibinfo
  {pages} {034612} (\bibinfo {year} {2019})}\BibitemShut {NoStop}%
\bibitem [{\citenamefont {Carjan}\ \emph {et~al.}(2019)\citenamefont {Carjan},
  \citenamefont {Ivanyuk},\ and\ \citenamefont {Oganessian}}]{Carjan2019}%
  \BibitemOpen
  \bibfield  {author} {\bibinfo {author} {\bibfnamefont {N.}~\bibnamefont
  {Carjan}}, \bibinfo {author} {\bibfnamefont {F.~A.}\ \bibnamefont {Ivanyuk}},
  \ and\ \bibinfo {author} {\bibfnamefont {Y.~T.}\ \bibnamefont {Oganessian}},\
  }\bibfield  {title} {\enquote {\bibinfo {title} {Fission of superheavy
  nuclei: Fragment mass distributions and their dependence on excitation
  energy},}\ }\href {\doibase 10.1103/PhysRevC.99.064606} {\bibfield  {journal}
  {\bibinfo  {journal} {Phys. Rev. C}\ }\textbf {\bibinfo {volume} {99}},\
  \bibinfo {pages} {064606} (\bibinfo {year} {2019})}\BibitemShut {NoStop}%
\bibitem [{\citenamefont {Pa\ifmmode~\mbox{\c{s}}\else \c{s}\fi{}ca}\ \emph
  {et~al.}(2019)\citenamefont {Pa\ifmmode~\mbox{\c{s}}\else \c{s}\fi{}ca},
  \citenamefont {Andreev}, \citenamefont {Adamian},\ and\ \citenamefont
  {Antonenko}}]{Pasca2019}%
  \BibitemOpen
  \bibfield  {author} {\bibinfo {author} {\bibfnamefont {H.}~\bibnamefont
  {Pa\ifmmode~\mbox{\c{s}}\else \c{s}\fi{}ca}}, \bibinfo {author}
  {\bibfnamefont {A.~V.}\ \bibnamefont {Andreev}}, \bibinfo {author}
  {\bibfnamefont {G.~G.}\ \bibnamefont {Adamian}}, \ and\ \bibinfo {author}
  {\bibfnamefont {N.~V.}\ \bibnamefont {Antonenko}},\ }\bibfield  {title}
  {\enquote {\bibinfo {title} {Change of the shape of mass and charge
  distributions in fission of {Cf} isotopes with excitation energy},}\ }\href
  {\doibase 10.1103/PhysRevC.99.064611} {\bibfield  {journal} {\bibinfo
  {journal} {Phys. Rev. C}\ }\textbf {\bibinfo {volume} {99}},\ \bibinfo
  {pages} {064611} (\bibinfo {year} {2019})}\BibitemShut {NoStop}%
\bibitem [{\citenamefont {Decowski}\ \emph {et~al.}(1968)\citenamefont
  {Decowski}, \citenamefont {Grochulski}, \citenamefont {Marcinkowski},
  \citenamefont {Siwek},\ and\ \citenamefont {Wilhelmi}}]{Dec68}%
  \BibitemOpen
  \bibfield  {author} {\bibinfo {author} {\bibfnamefont {P.}~\bibnamefont
  {Decowski}}, \bibinfo {author} {\bibfnamefont {W.}~\bibnamefont
  {Grochulski}}, \bibinfo {author} {\bibfnamefont {A.}~\bibnamefont
  {Marcinkowski}}, \bibinfo {author} {\bibfnamefont {K.}~\bibnamefont {Siwek}},
  \ and\ \bibinfo {author} {\bibfnamefont {Z.}~\bibnamefont {Wilhelmi}},\
  }\bibfield  {title} {\enquote {\bibinfo {title} {On superconductivity effects
  in nuclear level density},}\ }\href {\doibase
  https://doi.org/10.1016/0375-9474(68)90687-8} {\bibfield  {journal} {\bibinfo
   {journal} {Nucl. Phys. A}\ }\textbf {\bibinfo {volume} {110}},\ \bibinfo
  {pages} {129} (\bibinfo {year} {1968})}\BibitemShut {NoStop}%
\bibitem [{\citenamefont {Demetriou}\ and\ \citenamefont
  {Goriely}(2001)}]{Dem01}%
  \BibitemOpen
  \bibfield  {author} {\bibinfo {author} {\bibfnamefont {P.}~\bibnamefont
  {Demetriou}}\ and\ \bibinfo {author} {\bibfnamefont {S.}~\bibnamefont
  {Goriely}},\ }\bibfield  {title} {\enquote {\bibinfo {title} {Microscopic
  nuclear level densities for practical applications},}\ }\href {\doibase
  https://doi.org/10.1016/S0375-9474(01)01095-8} {\bibfield  {journal}
  {\bibinfo  {journal} {Nucl. Phys. A}\ }\textbf {\bibinfo {volume} {695}},\
  \bibinfo {pages} {95} (\bibinfo {year} {2001})}\BibitemShut {NoStop}%
\bibitem [{\citenamefont {Mumpower}\ \emph {et~al.}(2020)\citenamefont
  {Mumpower}, \citenamefont {Jaffke}, \citenamefont {Verriere},\ and\
  \citenamefont {Randrup}}]{Mumpower2020}%
  \BibitemOpen
  \bibfield  {author} {\bibinfo {author} {\bibfnamefont {M.~R.}\ \bibnamefont
  {Mumpower}}, \bibinfo {author} {\bibfnamefont {P.}~\bibnamefont {Jaffke}},
  \bibinfo {author} {\bibfnamefont {M.}~\bibnamefont {Verriere}}, \ and\
  \bibinfo {author} {\bibfnamefont {J.}~\bibnamefont {Randrup}},\ }\bibfield
  {title} {\enquote {\bibinfo {title} {Primary fission fragment mass yields
  across the chart of nuclides},}\ }\href {\doibase
  10.1103/PhysRevC.101.054607} {\bibfield  {journal} {\bibinfo  {journal}
  {Phys. Rev. C}\ }\textbf {\bibinfo {volume} {101}},\ \bibinfo {pages}
  {054607} (\bibinfo {year} {2020})}\BibitemShut {NoStop}%
\bibitem [{\citenamefont {Albertsson}\ \emph {et~al.}(2020)\citenamefont
  {Albertsson}, \citenamefont {Carlsson}, \citenamefont {D{\o}ssing},
  \citenamefont {M{\"o}ller}, \citenamefont {Randrup},\ and\ \citenamefont
  {{\AA}berg}}]{Albertsson2020}%
  \BibitemOpen
  \bibfield  {author} {\bibinfo {author} {\bibfnamefont {M.}~\bibnamefont
  {Albertsson}}, \bibinfo {author} {\bibfnamefont {B.~G.}\ \bibnamefont
  {Carlsson}}, \bibinfo {author} {\bibfnamefont {T.}~\bibnamefont
  {D{\o}ssing}}, \bibinfo {author} {\bibfnamefont {P.}~\bibnamefont
  {M{\"o}ller}}, \bibinfo {author} {\bibfnamefont {J.}~\bibnamefont {Randrup}},
  \ and\ \bibinfo {author} {\bibfnamefont {S.}~\bibnamefont {{\AA}berg}},\
  }\bibfield  {title} {\enquote {\bibinfo {title} {Calculated fission-fragment
  mass yields and average total kinetic energies of heavy and superheavy
  nuclei},}\ }\href {\doibase 10.1140/epja/s10050-020-00036-9} {\bibfield
  {journal} {\bibinfo  {journal} {Eur. Phys. J. A}\ }\textbf {\bibinfo {volume}
  {56}},\ \bibinfo {pages} {46} (\bibinfo {year} {2020})}\BibitemShut {NoStop}%
\bibitem [{\citenamefont {Ishizuka}\ \emph {et~al.}(2017)\citenamefont
  {Ishizuka}, \citenamefont {Usang}, \citenamefont {Ivanyuk}, \citenamefont
  {Maruhn}, \citenamefont {Nishio},\ and\ \citenamefont
  {Chiba}}]{ishizuka2017}%
  \BibitemOpen
  \bibfield  {author} {\bibinfo {author} {\bibfnamefont {C.}~\bibnamefont
  {Ishizuka}}, \bibinfo {author} {\bibfnamefont {M.~D.}\ \bibnamefont {Usang}},
  \bibinfo {author} {\bibfnamefont {F.~A.}\ \bibnamefont {Ivanyuk}}, \bibinfo
  {author} {\bibfnamefont {J.~A.}\ \bibnamefont {Maruhn}}, \bibinfo {author}
  {\bibfnamefont {K.}~\bibnamefont {Nishio}}, \ and\ \bibinfo {author}
  {\bibfnamefont {S.}~\bibnamefont {Chiba}},\ }\bibfield  {title} {\enquote
  {\bibinfo {title} {Four-dimensional {L}angevin approach to low-energy nuclear
  fission of $^{236}${U}},}\ }\href {\doibase 10.1103/PhysRevC.96.064616}
  {\bibfield  {journal} {\bibinfo  {journal} {Phys. Rev. C}\ }\textbf {\bibinfo
  {volume} {96}},\ \bibinfo {pages} {064616} (\bibinfo {year}
  {2017})}\BibitemShut {NoStop}%
\bibitem [{\citenamefont {Sadhukhan}\ \emph {et~al.}(2016)\citenamefont
  {Sadhukhan}, \citenamefont {Nazarewicz},\ and\ \citenamefont
  {Schunck}}]{sadhukhan2016}%
  \BibitemOpen
  \bibfield  {author} {\bibinfo {author} {\bibfnamefont {J.}~\bibnamefont
  {Sadhukhan}}, \bibinfo {author} {\bibfnamefont {W.}~\bibnamefont
  {Nazarewicz}}, \ and\ \bibinfo {author} {\bibfnamefont {N.}~\bibnamefont
  {Schunck}},\ }\bibfield  {title} {\enquote {\bibinfo {title} {Microscopic
  modeling of mass and charge distributions in the spontaneous fission of
  $^{240}${Pu}},}\ }\href {\doibase 10.1103/PhysRevC.93.011304} {\bibfield
  {journal} {\bibinfo  {journal} {Phys. Rev. C}\ }\textbf {\bibinfo {volume}
  {93}},\ \bibinfo {pages} {011304} (\bibinfo {year} {2016})}\BibitemShut
  {NoStop}%
\bibitem [{\citenamefont {Bulgac}\ \emph {et~al.}(2019)\citenamefont {Bulgac},
  \citenamefont {Jin}, \citenamefont {Roche}, \citenamefont {Schunck},\ and\
  \citenamefont {Stetcu}}]{Bulgac2019a}%
  \BibitemOpen
  \bibfield  {author} {\bibinfo {author} {\bibfnamefont {A.}~\bibnamefont
  {Bulgac}}, \bibinfo {author} {\bibfnamefont {S.}~\bibnamefont {Jin}},
  \bibinfo {author} {\bibfnamefont {K.~J.}\ \bibnamefont {Roche}}, \bibinfo
  {author} {\bibfnamefont {N.}~\bibnamefont {Schunck}}, \ and\ \bibinfo
  {author} {\bibfnamefont {I.}~\bibnamefont {Stetcu}},\ }\bibfield  {title}
  {\enquote {\bibinfo {title} {Fission dynamics of $^{240}\mathrm{Pu}$ from
  saddle to scission and beyond},}\ }\href {\doibase
  10.1103/PhysRevC.100.034615} {\bibfield  {journal} {\bibinfo  {journal}
  {Phys. Rev. C}\ }\textbf {\bibinfo {volume} {100}},\ \bibinfo {pages}
  {034615} (\bibinfo {year} {2019})}\BibitemShut {NoStop}%
\bibitem [{\citenamefont {M{\"o}ller}\ and\ \citenamefont
  {Ichikawa}(2015)}]{moller2015c}%
  \BibitemOpen
  \bibfield  {author} {\bibinfo {author} {\bibfnamefont {P.}~\bibnamefont
  {M{\"o}ller}}\ and\ \bibinfo {author} {\bibfnamefont {T.}~\bibnamefont
  {Ichikawa}},\ }\bibfield  {title} {\enquote {\bibinfo {title} {A method to
  calculate fission-fragment yields {Y(Z,N)} versus proton and neutron number
  in the {B}rownian shape-motion model},}\ }\href {\doibase
  10.1140/epja/i2015-15173-1} {\bibfield  {journal} {\bibinfo  {journal} {Eur.
  Phys. J. A}\ }\textbf {\bibinfo {volume} {51}},\ \bibinfo {pages} {173}
  (\bibinfo {year} {2015})}\BibitemShut {NoStop}%
\bibitem [{\citenamefont {Verriere}\ and\ \citenamefont
  {Mumpower}(2021)}]{Verriere2021b}%
  \BibitemOpen
  \bibfield  {author} {\bibinfo {author} {\bibfnamefont {M.}~\bibnamefont
  {Verriere}}\ and\ \bibinfo {author} {\bibfnamefont {M.~R.}\ \bibnamefont
  {Mumpower}},\ }\bibfield  {title} {\enquote {\bibinfo {title} {Improvements
  to the macroscopic-microscopic approach of nuclear fission},}\ }\href
  {\doibase 10.1103/PhysRevC.103.034617} {\bibfield  {journal} {\bibinfo
  {journal} {Phys. Rev. C}\ }\textbf {\bibinfo {volume} {103}},\ \bibinfo
  {pages} {034617} (\bibinfo {year} {2021})}\BibitemShut {NoStop}%
\bibitem [{\citenamefont {Goddard}\ \emph {et~al.}(2015)\citenamefont
  {Goddard}, \citenamefont {Stevenson},\ and\ \citenamefont
  {Rios}}]{goddard2015}%
  \BibitemOpen
  \bibfield  {author} {\bibinfo {author} {\bibfnamefont {P.}~\bibnamefont
  {Goddard}}, \bibinfo {author} {\bibfnamefont {P.}~\bibnamefont {Stevenson}},
  \ and\ \bibinfo {author} {\bibfnamefont {A.}~\bibnamefont {Rios}},\
  }\bibfield  {title} {\enquote {\bibinfo {title} {Fission dynamics within
  time-dependent {Hartree-Fock: Deformation}-induced fission},}\ }\href
  {\doibase 10.1103/PhysRevC.92.054610} {\bibfield  {journal} {\bibinfo
  {journal} {Phys. Rev. C}\ }\textbf {\bibinfo {volume} {92}},\ \bibinfo
  {pages} {054610} (\bibinfo {year} {2015})}\BibitemShut {NoStop}%
\bibitem [{\citenamefont {Tanimura}\ \emph {et~al.}(2017)\citenamefont
  {Tanimura}, \citenamefont {Lacroix},\ and\ \citenamefont
  {Ayik}}]{tanimura2017}%
  \BibitemOpen
  \bibfield  {author} {\bibinfo {author} {\bibfnamefont {Y.}~\bibnamefont
  {Tanimura}}, \bibinfo {author} {\bibfnamefont {D.}~\bibnamefont {Lacroix}}, \
  and\ \bibinfo {author} {\bibfnamefont {S.}~\bibnamefont {Ayik}},\ }\bibfield
  {title} {\enquote {\bibinfo {title} {Microscopic phase-space exploration
  modeling of $^{258}${F}m spontaneous fission},}\ }\href {\doibase
  10.1103/PhysRevLett.118.152501} {\bibfield  {journal} {\bibinfo  {journal}
  {Phys. Rev. Lett.}\ }\textbf {\bibinfo {volume} {118}},\ \bibinfo {pages}
  {152501} (\bibinfo {year} {2017})}\BibitemShut {NoStop}%
\bibitem [{\citenamefont {Scamps}\ and\ \citenamefont
  {Simenel}(2018)}]{scamps2018a}%
  \BibitemOpen
  \bibfield  {author} {\bibinfo {author} {\bibfnamefont {G.}~\bibnamefont
  {Scamps}}\ and\ \bibinfo {author} {\bibfnamefont {C.}~\bibnamefont
  {Simenel}},\ }\bibfield  {title} {\enquote {\bibinfo {title} {Impact of
  pear-shaped fission fragments on mass-asymmetric fission in actinides},}\
  }\href {\doibase 10.1038/s41586-018-0780-0} {\bibfield  {journal} {\bibinfo
  {journal} {Nature}\ }\textbf {\bibinfo {volume} {564}},\ \bibinfo {pages}
  {382} (\bibinfo {year} {2018})}\BibitemShut {NoStop}%
\bibitem [{\citenamefont {Scamps}\ and\ \citenamefont
  {Simenel}(2019)}]{scamps2019a}%
  \BibitemOpen
  \bibfield  {author} {\bibinfo {author} {\bibfnamefont {G.}~\bibnamefont
  {Scamps}}\ and\ \bibinfo {author} {\bibfnamefont {C.}~\bibnamefont
  {Simenel}},\ }\bibfield  {title} {\enquote {\bibinfo {title} {Effect of shell
  structure on the fission of sub-lead nuclei},}\ }\href {\doibase
  10.1103/PhysRevC.100.041602} {\bibfield  {journal} {\bibinfo  {journal}
  {Phys. Rev. C}\ }\textbf {\bibinfo {volume} {100}},\ \bibinfo {pages}
  {041602} (\bibinfo {year} {2019})}\BibitemShut {NoStop}%
\bibitem [{\citenamefont {Regnier}\ \emph {et~al.}(2019)\citenamefont
  {Regnier}, \citenamefont {Dubray},\ and\ \citenamefont
  {Schunck}}]{regnier2019}%
  \BibitemOpen
  \bibfield  {author} {\bibinfo {author} {\bibfnamefont {D.}~\bibnamefont
  {Regnier}}, \bibinfo {author} {\bibfnamefont {N.}~\bibnamefont {Dubray}}, \
  and\ \bibinfo {author} {\bibfnamefont {N.}~\bibnamefont {Schunck}},\
  }\bibfield  {title} {\enquote {\bibinfo {title} {{From asymmetric to
  symmetric fission in the fermium isotopes within the time-dependent
  generator-coordinate-method formalism}},}\ }\href {\doibase
  10.1103/PhysRevC.99.024611} {\bibfield  {journal} {\bibinfo  {journal} {Phys.
  Rev. C}\ }\textbf {\bibinfo {volume} {99}},\ \bibinfo {pages} {024611}
  (\bibinfo {year} {2019})}\BibitemShut {NoStop}%
\bibitem [{\citenamefont {Zhao}\ \emph {et~al.}(2020)\citenamefont {Zhao},
  \citenamefont {Nik\ifmmode \check{s}\else \v{s}\fi{}i\ifmmode~\acute{c}\else
  \'{c}\fi{}}, \citenamefont {Vretenar},\ and\ \citenamefont
  {Zhou}}]{zhao2020}%
  \BibitemOpen
  \bibfield  {author} {\bibinfo {author} {\bibfnamefont {J.}~\bibnamefont
  {Zhao}}, \bibinfo {author} {\bibfnamefont {T.}~\bibnamefont {Nik\ifmmode
  \check{s}\else \v{s}\fi{}i\ifmmode~\acute{c}\else \'{c}\fi{}}}, \bibinfo
  {author} {\bibfnamefont {D.}~\bibnamefont {Vretenar}}, \ and\ \bibinfo
  {author} {\bibfnamefont {S.-G.}\ \bibnamefont {Zhou}},\ }\bibfield  {title}
  {\enquote {\bibinfo {title} {Time-dependent generator coordinate method study
  of fission: Mass parameters},}\ }\href {\doibase 10.1103/PhysRevC.101.064605}
  {\bibfield  {journal} {\bibinfo  {journal} {Phys. Rev. C}\ }\textbf {\bibinfo
  {volume} {101}},\ \bibinfo {pages} {064605} (\bibinfo {year}
  {2020})}\BibitemShut {NoStop}%
\bibitem [{\citenamefont {Verriere}\ and\ \citenamefont
  {Regnier}(2020)}]{verriere2020}%
  \BibitemOpen
  \bibfield  {author} {\bibinfo {author} {\bibfnamefont {M.}~\bibnamefont
  {Verriere}}\ and\ \bibinfo {author} {\bibfnamefont {D.}~\bibnamefont
  {Regnier}},\ }\bibfield  {title} {\enquote {\bibinfo {title} {The
  time-dependent generator coordinate method in nuclear physics},}\ }\href
  {\doibase 10.3389/fphy.2020.00233} {\bibfield  {journal} {\bibinfo  {journal}
  {Front. Phys.}\ }\textbf {\bibinfo {volume} {8}},\ \bibinfo {pages} {233}
  (\bibinfo {year} {2020})}\BibitemShut {NoStop}%
\bibitem [{\citenamefont {Simenel}(2010)}]{Simenel2010}%
  \BibitemOpen
  \bibfield  {author} {\bibinfo {author} {\bibfnamefont {C.}~\bibnamefont
  {Simenel}},\ }\bibfield  {title} {\enquote {\bibinfo {title} {Particle
  transfer reactions with the time-dependent {Hartree-Fock} theory using a
  particle number projection technique},}\ }\href {\doibase
  10.1103/PhysRevLett.105.192701} {\bibfield  {journal} {\bibinfo  {journal}
  {Phys. Rev. Lett.}\ }\textbf {\bibinfo {volume} {105}},\ \bibinfo {pages}
  {192701} (\bibinfo {year} {2010})}\BibitemShut {NoStop}%
\bibitem [{\citenamefont {Scamps}\ \emph {et~al.}(2015)\citenamefont {Scamps},
  \citenamefont {Simenel},\ and\ \citenamefont {Lacroix}}]{scamps2015}%
  \BibitemOpen
  \bibfield  {author} {\bibinfo {author} {\bibfnamefont {G.}~\bibnamefont
  {Scamps}}, \bibinfo {author} {\bibfnamefont {C.}~\bibnamefont {Simenel}}, \
  and\ \bibinfo {author} {\bibfnamefont {D.}~\bibnamefont {Lacroix}},\
  }\bibfield  {title} {\enquote {\bibinfo {title} {Superfluid dynamics of
  $^{258}${Fm} fission},}\ }\href {\doibase 10.1103/PhysRevC.92.011602}
  {\bibfield  {journal} {\bibinfo  {journal} {Phys. Rev. C}\ }\textbf {\bibinfo
  {volume} {92}},\ \bibinfo {pages} {011602} (\bibinfo {year}
  {2015})}\BibitemShut {NoStop}%
\bibitem [{\citenamefont {Verriere}\ \emph {et~al.}(2019)\citenamefont
  {Verriere}, \citenamefont {Schunck},\ and\ \citenamefont
  {Kawano}}]{Verriere2019}%
  \BibitemOpen
  \bibfield  {author} {\bibinfo {author} {\bibfnamefont {M.}~\bibnamefont
  {Verriere}}, \bibinfo {author} {\bibfnamefont {N.}~\bibnamefont {Schunck}}, \
  and\ \bibinfo {author} {\bibfnamefont {T.}~\bibnamefont {Kawano}},\
  }\bibfield  {title} {\enquote {\bibinfo {title} {Number of particles in
  fission fragments},}\ }\href {\doibase 10.1103/PhysRevC.100.024612}
  {\bibfield  {journal} {\bibinfo  {journal} {Phys. Rev. C}\ }\textbf {\bibinfo
  {volume} {100}},\ \bibinfo {pages} {024612} (\bibinfo {year}
  {2019})}\BibitemShut {NoStop}%
\bibitem [{\citenamefont {Verriere}\ \emph {et~al.}(2021)\citenamefont
  {Verriere}, \citenamefont {Schunck},\ and\ \citenamefont
  {Regnier}}]{Verriere2021}%
  \BibitemOpen
  \bibfield  {author} {\bibinfo {author} {\bibfnamefont {M.}~\bibnamefont
  {Verriere}}, \bibinfo {author} {\bibfnamefont {N.}~\bibnamefont {Schunck}}, \
  and\ \bibinfo {author} {\bibfnamefont {D.}~\bibnamefont {Regnier}},\
  }\bibfield  {title} {\enquote {\bibinfo {title} {Microscopic calculation of
  fission product yields with particle-number projection},}\ }\href {\doibase
  10.1103/PhysRevC.103.054602} {\bibfield  {journal} {\bibinfo  {journal}
  {Phys. Rev. C}\ }\textbf {\bibinfo {volume} {103}},\ \bibinfo {pages}
  {054602} (\bibinfo {year} {2021})}\BibitemShut {NoStop}%
\bibitem [{\citenamefont {Sadhukhan}\ \emph {et~al.}(2020)\citenamefont
  {Sadhukhan}, \citenamefont {Giuliani}, \citenamefont {Matheson},\ and\
  \citenamefont {Nazarewicz}}]{Sadhukhan2020}%
  \BibitemOpen
  \bibfield  {author} {\bibinfo {author} {\bibfnamefont {J.}~\bibnamefont
  {Sadhukhan}}, \bibinfo {author} {\bibfnamefont {S.~A.}\ \bibnamefont
  {Giuliani}}, \bibinfo {author} {\bibfnamefont {Z.}~\bibnamefont {Matheson}},
  \ and\ \bibinfo {author} {\bibfnamefont {W.}~\bibnamefont {Nazarewicz}},\
  }\bibfield  {title} {\enquote {\bibinfo {title} {Efficient method for
  estimation of fission fragment yields of $r$-process nuclei},}\ }\href
  {\doibase 10.1103/PhysRevC.101.065803} {\bibfield  {journal} {\bibinfo
  {journal} {Phys. Rev. C}\ }\textbf {\bibinfo {volume} {101}},\ \bibinfo
  {pages} {065803} (\bibinfo {year} {2020})}\BibitemShut {NoStop}%
\bibitem [{\citenamefont {Zhang}\ \emph {et~al.}(2016)\citenamefont {Zhang},
  \citenamefont {Schuetrumpf},\ and\ \citenamefont {Nazarewicz}}]{zhang2016}%
  \BibitemOpen
  \bibfield  {author} {\bibinfo {author} {\bibfnamefont {C.~L.}\ \bibnamefont
  {Zhang}}, \bibinfo {author} {\bibfnamefont {B.}~\bibnamefont {Schuetrumpf}},
  \ and\ \bibinfo {author} {\bibfnamefont {W.}~\bibnamefont {Nazarewicz}},\
  }\bibfield  {title} {\enquote {\bibinfo {title} {Nucleon localization and
  fragment formation in nuclear fission},}\ }\href {\doibase
  10.1103/PhysRevC.94.064323} {\bibfield  {journal} {\bibinfo  {journal} {Phys.
  Rev. C}\ }\textbf {\bibinfo {volume} {94}},\ \bibinfo {pages} {064323}
  (\bibinfo {year} {2016})}\BibitemShut {NoStop}%
\bibitem [{\citenamefont {Caama{\~n}o}\ and\ \citenamefont
  {Farget}(2017)}]{Caamano2017}%
  \BibitemOpen
  \bibfield  {author} {\bibinfo {author} {\bibfnamefont {M.}~\bibnamefont
  {Caama{\~n}o}}\ and\ \bibinfo {author} {\bibfnamefont {F.}~\bibnamefont
  {Farget}},\ }\bibfield  {title} {\enquote {\bibinfo {title} {Energy balance
  and deformation at scission in $^{240}${Pu} fission},}\ }\href {\doibase
  https://doi.org/10.1016/j.physletb.2017.04.041} {\bibfield  {journal}
  {\bibinfo  {journal} {Phys. Lett. B}\ }\textbf {\bibinfo {volume} {770}},\
  \bibinfo {pages} {72} (\bibinfo {year} {2017})}\BibitemShut {NoStop}%
\bibitem [{\citenamefont {Matheson}\ \emph {et~al.}(2019)\citenamefont
  {Matheson}, \citenamefont {Giuliani}, \citenamefont {Nazarewicz},
  \citenamefont {Sadhukhan},\ and\ \citenamefont {Schunck}}]{matheson2019}%
  \BibitemOpen
  \bibfield  {author} {\bibinfo {author} {\bibfnamefont {Z.}~\bibnamefont
  {Matheson}}, \bibinfo {author} {\bibfnamefont {S.~A.}\ \bibnamefont
  {Giuliani}}, \bibinfo {author} {\bibfnamefont {W.}~\bibnamefont
  {Nazarewicz}}, \bibinfo {author} {\bibfnamefont {J.}~\bibnamefont
  {Sadhukhan}}, \ and\ \bibinfo {author} {\bibfnamefont {N.}~\bibnamefont
  {Schunck}},\ }\bibfield  {title} {\enquote {\bibinfo {title} {Cluster
  radioactivity of $^{294}${Og}},}\ }\href {\doibase
  10.1103/PhysRevC.99.041304} {\bibfield  {journal} {\bibinfo  {journal} {Phys.
  Rev. C}\ }\textbf {\bibinfo {volume} {99}},\ \bibinfo {pages} {041304}
  (\bibinfo {year} {2019})}\BibitemShut {NoStop}%
\bibitem [{\citenamefont {Pei}\ \emph {et~al.}(2009)\citenamefont {Pei},
  \citenamefont {Nazarewicz}, \citenamefont {Sheikh},\ and\ \citenamefont
  {Kerman}}]{pei2009}%
  \BibitemOpen
  \bibfield  {author} {\bibinfo {author} {\bibfnamefont {J.~C.}\ \bibnamefont
  {Pei}}, \bibinfo {author} {\bibfnamefont {W.}~\bibnamefont {Nazarewicz}},
  \bibinfo {author} {\bibfnamefont {J.~A.}\ \bibnamefont {Sheikh}}, \ and\
  \bibinfo {author} {\bibfnamefont {A.~K.}\ \bibnamefont {Kerman}},\ }\bibfield
   {title} {\enquote {\bibinfo {title} {Fission barriers of compound superheavy
  nuclei},}\ }\href {\doibase 10.1103/PhysRevLett.102.192501} {\bibfield
  {journal} {\bibinfo  {journal} {Phys. Rev. Lett.}\ }\textbf {\bibinfo
  {volume} {102}},\ \bibinfo {pages} {192501} (\bibinfo {year}
  {2009})}\BibitemShut {NoStop}%
\bibitem [{\citenamefont {Schunck}\ \emph {et~al.}(2017)\citenamefont
  {Schunck}, \citenamefont {Dobaczewski}, \citenamefont {Satu{\l}a},
  \citenamefont {B{\c{a}}czyk}, \citenamefont {Dudek}, \citenamefont {Gao},
  \citenamefont {Konieczka}, \citenamefont {Sato}, \citenamefont {Shi},
  \citenamefont {Wang},\ and\ \citenamefont {Werner}}]{schunck2017}%
  \BibitemOpen
  \bibfield  {author} {\bibinfo {author} {\bibfnamefont {N.}~\bibnamefont
  {Schunck}}, \bibinfo {author} {\bibfnamefont {J.}~\bibnamefont
  {Dobaczewski}}, \bibinfo {author} {\bibfnamefont {W.}~\bibnamefont
  {Satu{\l}a}}, \bibinfo {author} {\bibfnamefont {P.}~\bibnamefont
  {B{\c{a}}czyk}}, \bibinfo {author} {\bibfnamefont {J.}~\bibnamefont {Dudek}},
  \bibinfo {author} {\bibfnamefont {Y.}~\bibnamefont {Gao}}, \bibinfo {author}
  {\bibfnamefont {M.}~\bibnamefont {Konieczka}}, \bibinfo {author}
  {\bibfnamefont {K.}~\bibnamefont {Sato}}, \bibinfo {author} {\bibfnamefont
  {Y.}~\bibnamefont {Shi}}, \bibinfo {author} {\bibfnamefont {X.}~\bibnamefont
  {Wang}}, \ and\ \bibinfo {author} {\bibfnamefont {T.}~\bibnamefont
  {Werner}},\ }\bibfield  {title} {\enquote {\bibinfo {title} {{Solution of the
  Skyrme-Hartree-Fock-Bogolyubov equations in the Cartesian deformed
  harmonic-oscillator basis. (VIII) HFODD (v2.73y): A new version of the
  program}},}\ }\href {\doibase 10.1016/j.cpc.2017.03.007} {\bibfield
  {journal} {\bibinfo  {journal} {Comput. Phys. Commun.}\ }\textbf {\bibinfo
  {volume} {216}},\ \bibinfo {pages} {145} (\bibinfo {year}
  {2017})}\BibitemShut {NoStop}%
\bibitem [{\citenamefont {Bartel}\ \emph {et~al.}(1982)\citenamefont {Bartel},
  \citenamefont {Quentin}, \citenamefont {Brack}, \citenamefont {Guet},\ and\
  \citenamefont {H{\aa}kansson}}]{bartel1982}%
  \BibitemOpen
  \bibfield  {author} {\bibinfo {author} {\bibfnamefont {J.}~\bibnamefont
  {Bartel}}, \bibinfo {author} {\bibfnamefont {P.}~\bibnamefont {Quentin}},
  \bibinfo {author} {\bibfnamefont {M.}~\bibnamefont {Brack}}, \bibinfo
  {author} {\bibfnamefont {C.}~\bibnamefont {Guet}}, \ and\ \bibinfo {author}
  {\bibfnamefont {H.-B.}\ \bibnamefont {H{\aa}kansson}},\ }\bibfield  {title}
  {\enquote {\bibinfo {title} {{Towards a better parametrisation of Skyrme-like
  effective forces: A critical study of the SkM force}},}\ }\href {\doibase
  10.1016/0375-9474(82)90403-1} {\bibfield  {journal} {\bibinfo  {journal}
  {Nucl. Phys. A}\ }\textbf {\bibinfo {volume} {386}},\ \bibinfo {pages} {79}
  (\bibinfo {year} {1982})}\BibitemShut {NoStop}%
\bibitem [{\citenamefont {Dobaczewski}\ \emph {et~al.}(2002)\citenamefont
  {Dobaczewski}, \citenamefont {Nazarewicz},\ and\ \citenamefont
  {Stoitsov}}]{dobaczewski2002}%
  \BibitemOpen
  \bibfield  {author} {\bibinfo {author} {\bibfnamefont {J.}~\bibnamefont
  {Dobaczewski}}, \bibinfo {author} {\bibfnamefont {W.}~\bibnamefont
  {Nazarewicz}}, \ and\ \bibinfo {author} {\bibfnamefont {M.}~\bibnamefont
  {Stoitsov}},\ }\bibfield  {title} {\enquote {\bibinfo {title} {Nuclear
  ground-state properties from mean-field calculations},}\ }\href {\doibase
  10.1140/epja/i2001-10218-8} {\bibfield  {journal} {\bibinfo  {journal} {Eur.
  Phys. J. A}\ }\textbf {\bibinfo {volume} {15}},\ \bibinfo {pages} {21}
  (\bibinfo {year} {2002})}\BibitemShut {NoStop}%
\bibitem [{\citenamefont {Fong}(1953)}]{fong1953}%
  \BibitemOpen
  \bibfield  {author} {\bibinfo {author} {\bibfnamefont {P.}~\bibnamefont
  {Fong}},\ }\bibfield  {title} {\enquote {\bibinfo {title} {Asymmetric
  fission},}\ }\href {\doibase 10.1103/PhysRev.89.332} {\bibfield  {journal}
  {\bibinfo  {journal} {Phys. Rev.}\ }\textbf {\bibinfo {volume} {89}},\
  \bibinfo {pages} {332} (\bibinfo {year} {1953})}\BibitemShut {NoStop}%
\bibitem [{\citenamefont {Bondorf}\ \emph {et~al.}(1995)\citenamefont
  {Bondorf}, \citenamefont {Botvina}, \citenamefont {Iljinov}, \citenamefont
  {Mishustin},\ and\ \citenamefont {Sneppen}}]{Bondorf1995}%
  \BibitemOpen
  \bibfield  {author} {\bibinfo {author} {\bibfnamefont {J.}~\bibnamefont
  {Bondorf}}, \bibinfo {author} {\bibfnamefont {A.}~\bibnamefont {Botvina}},
  \bibinfo {author} {\bibfnamefont {A.}~\bibnamefont {Iljinov}}, \bibinfo
  {author} {\bibfnamefont {I.}~\bibnamefont {Mishustin}}, \ and\ \bibinfo
  {author} {\bibfnamefont {K.}~\bibnamefont {Sneppen}},\ }\bibfield  {title}
  {\enquote {\bibinfo {title} {Statistical multifragmentation of nuclei},}\
  }\href {\doibase 10.1016/0370-1573(94)00097-M} {\bibfield  {journal}
  {\bibinfo  {journal} {Phys. Rep.}\ }\textbf {\bibinfo {volume} {257}},\
  \bibinfo {pages} {133} (\bibinfo {year} {1995})}\BibitemShut {NoStop}%
\bibitem [{\citenamefont {Fong}(1956)}]{Fong56}%
  \BibitemOpen
  \bibfield  {author} {\bibinfo {author} {\bibfnamefont {P.}~\bibnamefont
  {Fong}},\ }\bibfield  {title} {\enquote {\bibinfo {title} {Statistical theory
  of nuclear fission: Asymmetric fission},}\ }\href {\doibase
  10.1103/PhysRev.102.434} {\bibfield  {journal} {\bibinfo  {journal} {Phys.
  Rev.}\ }\textbf {\bibinfo {volume} {102}},\ \bibinfo {pages} {434--448}
  (\bibinfo {year} {1956})}\BibitemShut {NoStop}%
\bibitem [{\citenamefont {Myers}\ and\ \citenamefont
  {Swiatecki}(1966)}]{myers1966}%
  \BibitemOpen
  \bibfield  {author} {\bibinfo {author} {\bibfnamefont {W.~D.}\ \bibnamefont
  {Myers}}\ and\ \bibinfo {author} {\bibfnamefont {W.~J.}\ \bibnamefont
  {Swiatecki}},\ }\bibfield  {title} {\enquote {\bibinfo {title} {Nuclear
  masses and deformations},}\ }\href {\doibase 10.1016/0029-5582(66)90639-0}
  {\bibfield  {journal} {\bibinfo  {journal} {Nucl. Phys.}\ }\textbf {\bibinfo
  {volume} {81}},\ \bibinfo {pages} {1} (\bibinfo {year} {1966})}\BibitemShut
  {NoStop}%
\bibitem [{\citenamefont {Wong}(1973)}]{won73}%
  \BibitemOpen
  \bibfield  {author} {\bibinfo {author} {\bibfnamefont {C.~Y.}\ \bibnamefont
  {Wong}},\ }\bibfield  {title} {\enquote {\bibinfo {title} {Interaction
  barrier in charged-particle nuclear reactions},}\ }\href {\doibase
  10.1103/PhysRevLett.31.766} {\bibfield  {journal} {\bibinfo  {journal} {Phys.
  Rev. Lett.}\ }\textbf {\bibinfo {volume} {31}},\ \bibinfo {pages} {766}
  (\bibinfo {year} {1973})}\BibitemShut {NoStop}%
\bibitem [{\citenamefont {Bertsch}\ \emph {et~al.}(2009)\citenamefont
  {Bertsch}, \citenamefont {Bertulani}, \citenamefont {Nazarewicz},
  \citenamefont {Schunck},\ and\ \citenamefont {Stoitsov}}]{bertsch2009}%
  \BibitemOpen
  \bibfield  {author} {\bibinfo {author} {\bibfnamefont {G.~F.}\ \bibnamefont
  {Bertsch}}, \bibinfo {author} {\bibfnamefont {C.~A.}\ \bibnamefont
  {Bertulani}}, \bibinfo {author} {\bibfnamefont {W.}~\bibnamefont
  {Nazarewicz}}, \bibinfo {author} {\bibfnamefont {N.}~\bibnamefont {Schunck}},
  \ and\ \bibinfo {author} {\bibfnamefont {M.~V.}\ \bibnamefont {Stoitsov}},\
  }\bibfield  {title} {\enquote {\bibinfo {title} {{Odd-even mass differences
  from self-consistent mean field theory}},}\ }\href {\doibase
  10.1103/PhysRevC.79.034306} {\bibfield  {journal} {\bibinfo  {journal} {Phys.
  Rev. C}\ }\textbf {\bibinfo {volume} {79}},\ \bibinfo {pages} {034306}
  (\bibinfo {year} {2009})}\BibitemShut {NoStop}%
\bibitem [{\citenamefont {Regnier}\ \emph {et~al.}(2016)\citenamefont
  {Regnier}, \citenamefont {Dubray}, \citenamefont {Schunck},\ and\
  \citenamefont {Verri{\`{e}}re}}]{regnier2016}%
  \BibitemOpen
  \bibfield  {author} {\bibinfo {author} {\bibfnamefont {D.}~\bibnamefont
  {Regnier}}, \bibinfo {author} {\bibfnamefont {N.}~\bibnamefont {Dubray}},
  \bibinfo {author} {\bibfnamefont {N.}~\bibnamefont {Schunck}}, \ and\
  \bibinfo {author} {\bibfnamefont {M.}~\bibnamefont {Verri{\`{e}}re}},\
  }\bibfield  {title} {\enquote {\bibinfo {title} {Fission fragment charge and
  mass distributions in $^{239}\mathrm{Pu}(n,f)$ in the adiabatic nuclear
  energy density functional theory},}\ }\href {\doibase
  10.1103/PhysRevC.93.054611} {\bibfield  {journal} {\bibinfo  {journal} {Phys.
  Rev. C}\ }\textbf {\bibinfo {volume} {93}},\ \bibinfo {pages} {054611}
  (\bibinfo {year} {2016})}\BibitemShut {NoStop}%
\bibitem [{\citenamefont {Lema{\^{i}}tre}\ \emph {et~al.}(2021)\citenamefont
  {Lema{\^{i}}tre}, \citenamefont {Goriely}, \citenamefont {Bauswein},\ and\
  \citenamefont {Janka}}]{lemaitre2021}%
  \BibitemOpen
  \bibfield  {author} {\bibinfo {author} {\bibfnamefont {J.-F.}\ \bibnamefont
  {Lema{\^{i}}tre}}, \bibinfo {author} {\bibfnamefont {S.}~\bibnamefont
  {Goriely}}, \bibinfo {author} {\bibfnamefont {A.}~\bibnamefont {Bauswein}}, \
  and\ \bibinfo {author} {\bibfnamefont {H.-T.}\ \bibnamefont {Janka}},\
  }\bibfield  {title} {\enquote {\bibinfo {title} {{Fission fragment
  distributions and their impact on the r-process nucleosynthesis in neutron
  star mergers}},}\ }\href {\doibase 10.1103/PhysRevC.103.025806} {\bibfield
  {journal} {\bibinfo  {journal} {Phys. Rev. C}\ }\textbf {\bibinfo {volume}
  {103}},\ \bibinfo {pages} {025806} (\bibinfo {year} {2021})}\BibitemShut
  {NoStop}%
\bibitem [{mas()}]{massexplorer}%
  \BibitemOpen
  \href@noop {} {}\bibinfo {note} {MassExplorer database:
  \url{http://massexplorer.frib.msu.edu/}}\BibitemShut {NoStop}%
\bibitem [{\citenamefont {Laidler}\ and\ \citenamefont
  {Brown}(1962)}]{laidler1962}%
  \BibitemOpen
  \bibfield  {author} {\bibinfo {author} {\bibfnamefont {J.}~\bibnamefont
  {Laidler}}\ and\ \bibinfo {author} {\bibfnamefont {F.}~\bibnamefont
  {Brown}},\ }\bibfield  {title} {\enquote {\bibinfo {title} {Mass distribution
  in the spontaneous fission of $^{240}${Pu}},}\ }\href {\doibase
  10.1016/0022-1902(62)80001-3} {\bibfield  {journal} {\bibinfo  {journal} {J.
  Inorg. Nucl. Chem.}\ }\textbf {\bibinfo {volume} {24}},\ \bibinfo {pages}
  {1485} (\bibinfo {year} {1962})}\BibitemShut {NoStop}%
\bibitem [{\citenamefont {Huang}\ \emph {et~al.}(2021)\citenamefont {Huang},
  \citenamefont {Wang}, \citenamefont {Kondev}, \citenamefont {Audi},\ and\
  \citenamefont {Naimi}}]{AME20}%
  \BibitemOpen
  \bibfield  {author} {\bibinfo {author} {\bibfnamefont {W.}~\bibnamefont
  {Huang}}, \bibinfo {author} {\bibfnamefont {M.}~\bibnamefont {Wang}},
  \bibinfo {author} {\bibfnamefont {F.}~\bibnamefont {Kondev}}, \bibinfo
  {author} {\bibfnamefont {G.}~\bibnamefont {Audi}}, \ and\ \bibinfo {author}
  {\bibfnamefont {S.}~\bibnamefont {Naimi}},\ }\bibfield  {title} {\enquote
  {\bibinfo {title} {The {AME} 2020 atomic mass evaluation {(I). Evaluation} of
  input data, and adjustment procedures},}\ }\href {\doibase
  10.1088/1674-1137/abddb0} {\bibfield  {journal} {\bibinfo  {journal} {Chin.
  Phys. C}\ }\textbf {\bibinfo {volume} {45}},\ \bibinfo {pages} {030002}
  (\bibinfo {year} {2021})}\BibitemShut {NoStop}%
\bibitem [{\citenamefont {Koning}\ \emph {et~al.}(2008)\citenamefont {Koning},
  \citenamefont {Hilaire},\ and\ \citenamefont {Duijvestijn}}]{koning2007}%
  \BibitemOpen
  \bibfield  {author} {\bibinfo {author} {\bibfnamefont {A.~J.}\ \bibnamefont
  {Koning}}, \bibinfo {author} {\bibfnamefont {S.}~\bibnamefont {Hilaire}}, \
  and\ \bibinfo {author} {\bibfnamefont {M.~C.}\ \bibnamefont {Duijvestijn}},\
  }\bibfield  {title} {\enquote {\bibinfo {title} {{TALYS}-1.0},}\ }\href
  {\doibase 10.1051/ndata:07767} {\bibfield  {journal} {\bibinfo  {journal}
  {ND2007 - International Conference on Nuclear Data for Science and
  Technology}\ ,\ \bibinfo {pages} {211--214}} (\bibinfo {year}
  {2008})}\BibitemShut {NoStop}%
\bibitem [{\citenamefont {Schmidt}\ \emph {et~al.}(2016)\citenamefont
  {Schmidt}, \citenamefont {Jurado}, \citenamefont {Amouroux},\ and\
  \citenamefont {Schmitt}}]{schmidt2016}%
  \BibitemOpen
  \bibfield  {author} {\bibinfo {author} {\bibfnamefont {K.-H.}\ \bibnamefont
  {Schmidt}}, \bibinfo {author} {\bibfnamefont {B.}~\bibnamefont {Jurado}},
  \bibinfo {author} {\bibfnamefont {C.}~\bibnamefont {Amouroux}}, \ and\
  \bibinfo {author} {\bibfnamefont {C.}~\bibnamefont {Schmitt}},\ }\bibfield
  {title} {\enquote {\bibinfo {title} {General description of fission
  observables: {GEF} model code},}\ }\href {\doibase 10.1016/j.nds.2015.12.009}
  {\bibfield  {journal} {\bibinfo  {journal} {Nucl. Data Sheets}\ }\textbf
  {\bibinfo {volume} {131}},\ \bibinfo {pages} {107--221} (\bibinfo {year}
  {2016})}\BibitemShut {NoStop}%
\bibitem [{\citenamefont {Kornilov}\ \emph {et~al.}(2003)\citenamefont
  {Kornilov}, \citenamefont {Kagalenko}, \citenamefont {Maslov},\ and\
  \citenamefont {Porodzinskij}}]{kor03}%
  \BibitemOpen
  \bibfield  {author} {\bibinfo {author} {\bibfnamefont {N.}~\bibnamefont
  {Kornilov}}, \bibinfo {author} {\bibfnamefont {A.}~\bibnamefont {Kagalenko}},
  \bibinfo {author} {\bibfnamefont {V.}~\bibnamefont {Maslov}}, \ and\ \bibinfo
  {author} {\bibfnamefont {Y.~V.}\ \bibnamefont {Porodzinskij}},\ }\bibfield
  {title} {\enquote {\bibinfo {title} {Neutron multiplicity for neutron
  incident energy from 0 to 150 {MeV}},}\ }\href
  {https://www.osti.gov/etdeweb/servlets/purl/20528880} {\bibfield  {journal}
  {\bibinfo  {journal} {Nuclear data section, INDC (CCP), IAEA}\ }\textbf
  {\bibinfo {volume} {437}},\ \bibinfo {pages} {1} (\bibinfo {year}
  {2003})}\BibitemShut {NoStop}%
\bibitem [{\citenamefont {Sh.~Zeynalov}(2011)}]{zeynalov2011}%
  \BibitemOpen
  \bibfield  {author} {\bibinfo {author} {\bibfnamefont {S.~O.}\ \bibnamefont
  {Sh.~Zeynalov}, \bibfnamefont {F.-J.~Hambsch}},\ }\bibfield  {title}
  {\enquote {\bibinfo {title} {Neutron emission in fission of
  $^{252}${Cf(SF)}},}\ }\href {\doibase 10.3938/jkps.59.1396} {\bibfield
  {journal} {\bibinfo  {journal} {J. Korean Phy. Soc.}\ }\textbf {\bibinfo
  {volume} {59}},\ \bibinfo {pages} {1396} (\bibinfo {year}
  {2011})}\BibitemShut {NoStop}%
\bibitem [{\citenamefont {Mariolopoulos}\ \emph
  {et~al.}(1981{\natexlab{b}})\citenamefont {Mariolopoulos}, \citenamefont
  {Hamelin}, \citenamefont {Blachot}, \citenamefont {Bocquet}, \citenamefont
  {Brissot}, \citenamefont {Crançon}, \citenamefont {Nifenecker},\ and\
  \citenamefont {Ristori}}]{mario1981}%
  \BibitemOpen
  \bibfield  {author} {\bibinfo {author} {\bibfnamefont {G.}~\bibnamefont
  {Mariolopoulos}}, \bibinfo {author} {\bibfnamefont {C.}~\bibnamefont
  {Hamelin}}, \bibinfo {author} {\bibfnamefont {J.}~\bibnamefont {Blachot}},
  \bibinfo {author} {\bibfnamefont {J.}~\bibnamefont {Bocquet}}, \bibinfo
  {author} {\bibfnamefont {R.}~\bibnamefont {Brissot}}, \bibinfo {author}
  {\bibfnamefont {J.}~\bibnamefont {Crançon}}, \bibinfo {author}
  {\bibfnamefont {H.}~\bibnamefont {Nifenecker}}, \ and\ \bibinfo {author}
  {\bibfnamefont {C.}~\bibnamefont {Ristori}},\ }\bibfield  {title} {\enquote
  {\bibinfo {title} {Charge distributions in low-energy nuclear fission and
  their relevance to fission dynamics},}\ }\href {\doibase
  10.1016/0375-9474(81)90477-2} {\bibfield  {journal} {\bibinfo  {journal}
  {Nucl. Phys. A}\ }\textbf {\bibinfo {volume} {361}},\ \bibinfo {pages} {213}
  (\bibinfo {year} {1981}{\natexlab{b}})}\BibitemShut {NoStop}%
\bibitem [{\citenamefont {Flynn}\ \emph {et~al.}(1975)\citenamefont {Flynn},
  \citenamefont {Gindler}, \citenamefont {Sjoblom},\ and\ \citenamefont
  {Glendenin}}]{flynn1975a}%
  \BibitemOpen
  \bibfield  {author} {\bibinfo {author} {\bibfnamefont {K.~F.}\ \bibnamefont
  {Flynn}}, \bibinfo {author} {\bibfnamefont {J.~E.}\ \bibnamefont {Gindler}},
  \bibinfo {author} {\bibfnamefont {R.~K.}\ \bibnamefont {Sjoblom}}, \ and\
  \bibinfo {author} {\bibfnamefont {L.~E.}\ \bibnamefont {Glendenin}},\
  }\bibfield  {title} {\enquote {\bibinfo {title} {Mass distributions for
  thermal-neutron-induced fission of $^{255}\mathrm{Fm}$ and
  $^{251}\mathrm{Cf}$},}\ }\href {\doibase 10.1103/PhysRevC.11.1676} {\bibfield
   {journal} {\bibinfo  {journal} {Phys. Rev. C}\ }\textbf {\bibinfo {volume}
  {11}},\ \bibinfo {pages} {1676} (\bibinfo {year} {1975})}\BibitemShut
  {NoStop}%
\bibitem [{\citenamefont {Harbour}\ \emph {et~al.}(1973)\citenamefont
  {Harbour}, \citenamefont {MacMurdo}, \citenamefont {Troutner},\ and\
  \citenamefont {Hoehn}}]{harbour1973}%
  \BibitemOpen
  \bibfield  {author} {\bibinfo {author} {\bibfnamefont {R.~M.}\ \bibnamefont
  {Harbour}}, \bibinfo {author} {\bibfnamefont {K.~W.}\ \bibnamefont
  {MacMurdo}}, \bibinfo {author} {\bibfnamefont {D.~E.}\ \bibnamefont
  {Troutner}}, \ and\ \bibinfo {author} {\bibfnamefont {M.~V.}\ \bibnamefont
  {Hoehn}},\ }\bibfield  {title} {\enquote {\bibinfo {title} {Mass and nuclear
  charge distributions from the spontaneous fission of $^{254}${Fm}},}\ }\href
  {\doibase 10.1103/PhysRevC.8.1488} {\bibfield  {journal} {\bibinfo  {journal}
  {Phys. Rev. C}\ }\textbf {\bibinfo {volume} {8}},\ \bibinfo {pages} {1488}
  (\bibinfo {year} {1973})}\BibitemShut {NoStop}%
\bibitem [{\citenamefont {Gindler}\ \emph {et~al.}(1977)\citenamefont
  {Gindler}, \citenamefont {Flynn}, \citenamefont {Glendenin},\ and\
  \citenamefont {Sjoblom}}]{gindler1977}%
  \BibitemOpen
  \bibfield  {author} {\bibinfo {author} {\bibfnamefont {J.~E.}\ \bibnamefont
  {Gindler}}, \bibinfo {author} {\bibfnamefont {K.~F.}\ \bibnamefont {Flynn}},
  \bibinfo {author} {\bibfnamefont {L.~E.}\ \bibnamefont {Glendenin}}, \ and\
  \bibinfo {author} {\bibfnamefont {R.~K.}\ \bibnamefont {Sjoblom}},\
  }\bibfield  {title} {\enquote {\bibinfo {title} {Distribution of mass,
  kinetic energy, and neutron yield in the spontaneous fission of
  $^{257}${Fm}},}\ }\href {\doibase 10.1103/PhysRevC.16.1483} {\bibfield
  {journal} {\bibinfo  {journal} {Phys. Rev. C}\ }\textbf {\bibinfo {volume}
  {16}},\ \bibinfo {pages} {1483} (\bibinfo {year} {1977})}\BibitemShut
  {NoStop}%
\bibitem [{\citenamefont {Flynn}\ \emph {et~al.}(1972)\citenamefont {Flynn},
  \citenamefont {Horwitz}, \citenamefont {Bloomquist}, \citenamefont {Barnes},
  \citenamefont {Sjoblom}, \citenamefont {Fields},\ and\ \citenamefont
  {Glendenin}}]{flynn1972}%
  \BibitemOpen
  \bibfield  {author} {\bibinfo {author} {\bibfnamefont {K.~F.}\ \bibnamefont
  {Flynn}}, \bibinfo {author} {\bibfnamefont {E.~P.}\ \bibnamefont {Horwitz}},
  \bibinfo {author} {\bibfnamefont {C.~A.~A.}\ \bibnamefont {Bloomquist}},
  \bibinfo {author} {\bibfnamefont {R.~F.}\ \bibnamefont {Barnes}}, \bibinfo
  {author} {\bibfnamefont {R.~K.}\ \bibnamefont {Sjoblom}}, \bibinfo {author}
  {\bibfnamefont {P.~R.}\ \bibnamefont {Fields}}, \ and\ \bibinfo {author}
  {\bibfnamefont {L.~E.}\ \bibnamefont {Glendenin}},\ }\bibfield  {title}
  {\enquote {\bibinfo {title} {Distribution of mass in the spontaneous fission
  of $^{256}${Fm}},}\ }\href {\doibase 10.1103/PhysRevC.5.1725} {\bibfield
  {journal} {\bibinfo  {journal} {Phys. Rev. C}\ }\textbf {\bibinfo {volume}
  {5}},\ \bibinfo {pages} {1725} (\bibinfo {year} {1972})}\BibitemShut
  {NoStop}%
\bibitem [{\citenamefont {Staszczak}\ \emph {et~al.}(2009)\citenamefont
  {Staszczak}, \citenamefont {Baran}, \citenamefont {Dobaczewski},\ and\
  \citenamefont {Nazarewicz}}]{staszczak2009}%
  \BibitemOpen
  \bibfield  {author} {\bibinfo {author} {\bibfnamefont {A.}~\bibnamefont
  {Staszczak}}, \bibinfo {author} {\bibfnamefont {A.}~\bibnamefont {Baran}},
  \bibinfo {author} {\bibfnamefont {J.}~\bibnamefont {Dobaczewski}}, \ and\
  \bibinfo {author} {\bibfnamefont {W.}~\bibnamefont {Nazarewicz}},\ }\bibfield
   {title} {\enquote {\bibinfo {title} {Microscopic description of complex
  nuclear decay: {Multimodal} fission},}\ }\href {\doibase
  10.1103/PhysRevC.80.014309} {\bibfield  {journal} {\bibinfo  {journal} {Phys.
  Rev. C}\ }\textbf {\bibinfo {volume} {80}},\ \bibinfo {pages} {014309}
  (\bibinfo {year} {2009})}\BibitemShut {NoStop}%
\bibitem [{\citenamefont {Bocquet}\ and\ \citenamefont
  {Brissot}(1989)}]{bocquet1989}%
  \BibitemOpen
  \bibfield  {author} {\bibinfo {author} {\bibfnamefont {J.}~\bibnamefont
  {Bocquet}}\ and\ \bibinfo {author} {\bibfnamefont {R.}~\bibnamefont
  {Brissot}},\ }\bibfield  {title} {\enquote {\bibinfo {title} {{Mass, energy
  and nuclear charge distribution of fission fragments}},}\ }\href {\doibase
  10.1016/0375-9474(89)90663-5} {\bibfield  {journal} {\bibinfo  {journal}
  {Nucl. Phys. A}\ }\textbf {\bibinfo {volume} {502}},\ \bibinfo {pages}
  {213--232} (\bibinfo {year} {1989})}\BibitemShut {NoStop}%
\bibitem [{\citenamefont {Schmitt}\ and\ \citenamefont
  {M{\"{o}}ller}(2021)}]{schmitt2021}%
  \BibitemOpen
  \bibfield  {author} {\bibinfo {author} {\bibfnamefont {C.}~\bibnamefont
  {Schmitt}}\ and\ \bibinfo {author} {\bibfnamefont {P.}~\bibnamefont
  {M{\"{o}}ller}},\ }\bibfield  {title} {\enquote {\bibinfo {title} {{On the
  isotopic composition of fission fragments}},}\ }\href {\doibase
  10.1016/j.physletb.2020.136017} {\bibfield  {journal} {\bibinfo  {journal}
  {Phys. Lett. B}\ }\textbf {\bibinfo {volume} {812}},\ \bibinfo {pages}
  {136017} (\bibinfo {year} {2021})}\BibitemShut {NoStop}%
\bibitem [{\citenamefont {Iyer}\ and\ \citenamefont
  {Ganguly}(1971)}]{iyer1971}%
  \BibitemOpen
  \bibfield  {author} {\bibinfo {author} {\bibfnamefont {M.~R.}\ \bibnamefont
  {Iyer}}\ and\ \bibinfo {author} {\bibfnamefont {A.~K.}\ \bibnamefont
  {Ganguly}},\ }\bibfield  {title} {\enquote {\bibinfo {title} {Nuclear charge
  distribution in fission fragments},}\ }\href {\doibase
  10.1103/PhysRevC.3.785} {\bibfield  {journal} {\bibinfo  {journal} {Phys.
  Rev. C}\ }\textbf {\bibinfo {volume} {3}},\ \bibinfo {pages} {785--797}
  (\bibinfo {year} {1971})}\BibitemShut {NoStop}%
\bibitem [{\citenamefont {Schmitt}\ \emph {et~al.}(1984)\citenamefont
  {Schmitt}, \citenamefont {Guessous}, \citenamefont {Bocquet}, \citenamefont
  {Clerc}, \citenamefont {Brissot}, \citenamefont {Engelhardt}, \citenamefont
  {Faust}, \citenamefont {Gönnenwein}, \citenamefont {Mutterer}, \citenamefont
  {Nifenecker}, \citenamefont {Pannicke}, \citenamefont {Ristori},\ and\
  \citenamefont {Theobald}}]{schmitt1984}%
  \BibitemOpen
  \bibfield  {author} {\bibinfo {author} {\bibfnamefont {C.}~\bibnamefont
  {Schmitt}}, \bibinfo {author} {\bibfnamefont {A.}~\bibnamefont {Guessous}},
  \bibinfo {author} {\bibfnamefont {J.}~\bibnamefont {Bocquet}}, \bibinfo
  {author} {\bibfnamefont {H.-G.}\ \bibnamefont {Clerc}}, \bibinfo {author}
  {\bibfnamefont {R.}~\bibnamefont {Brissot}}, \bibinfo {author} {\bibfnamefont
  {D.}~\bibnamefont {Engelhardt}}, \bibinfo {author} {\bibfnamefont
  {H.}~\bibnamefont {Faust}}, \bibinfo {author} {\bibfnamefont
  {F.}~\bibnamefont {Gönnenwein}}, \bibinfo {author} {\bibfnamefont
  {M.}~\bibnamefont {Mutterer}}, \bibinfo {author} {\bibfnamefont
  {H.}~\bibnamefont {Nifenecker}}, \bibinfo {author} {\bibfnamefont
  {J.}~\bibnamefont {Pannicke}}, \bibinfo {author} {\bibfnamefont
  {C.}~\bibnamefont {Ristori}}, \ and\ \bibinfo {author} {\bibfnamefont
  {J.}~\bibnamefont {Theobald}},\ }\bibfield  {title} {\enquote {\bibinfo
  {title} {Fission yields at different fission-product kinetic energies for
  thermal-neutron-induced fission of $^{239}${P}u},}\ }\href {\doibase
  10.1016/0375-9474(84)90191-X} {\bibfield  {journal} {\bibinfo  {journal}
  {Nucl. Phys. A}\ }\textbf {\bibinfo {volume} {430}},\ \bibinfo {pages} {21}
  (\bibinfo {year} {1984})}\BibitemShut {NoStop}%
\bibitem [{\citenamefont {Langanke}\ \emph {et~al.}(1996)\citenamefont
  {Langanke}, \citenamefont {Dean}, \citenamefont {Radha},\ and\ \citenamefont
  {Koonin}}]{Langanke1996}%
  \BibitemOpen
  \bibfield  {author} {\bibinfo {author} {\bibfnamefont {K.}~\bibnamefont
  {Langanke}}, \bibinfo {author} {\bibfnamefont {D.}~\bibnamefont {Dean}},
  \bibinfo {author} {\bibfnamefont {P.}~\bibnamefont {Radha}}, \ and\ \bibinfo
  {author} {\bibfnamefont {S.}~\bibnamefont {Koonin}},\ }\bibfield  {title}
  {\enquote {\bibinfo {title} {Temperature dependence of pair correlations in
  nuclei in the iron region},}\ }\href {\doibase 10.1016/0375-9474(96)00139-X}
  {\bibfield  {journal} {\bibinfo  {journal} {Nucl. Phys. A}\ }\textbf
  {\bibinfo {volume} {602}},\ \bibinfo {pages} {244} (\bibinfo {year}
  {1996})}\BibitemShut {NoStop}%
\bibitem [{\citenamefont {Kaneko}\ and\ \citenamefont
  {Hasegawa}(2004)}]{Kaneko2004}%
  \BibitemOpen
  \bibfield  {author} {\bibinfo {author} {\bibfnamefont {K.}~\bibnamefont
  {Kaneko}}\ and\ \bibinfo {author} {\bibfnamefont {M.}~\bibnamefont
  {Hasegawa}},\ }\bibfield  {title} {\enquote {\bibinfo {title} {Pairing
  transition of nuclei at finite temperature},}\ }\href {\doibase
  10.1016/j.nuclphysa.2004.05.001} {\bibfield  {journal} {\bibinfo  {journal}
  {Nucl. Phys. A}\ }\textbf {\bibinfo {volume} {740}},\ \bibinfo {pages} {95}
  (\bibinfo {year} {2004})}\BibitemShut {NoStop}%
\bibitem [{\citenamefont {De~Clercq}\ \emph {et~al.}(1976)\citenamefont
  {De~Clercq}, \citenamefont {Jacobs}, \citenamefont {De~Frenne}, \citenamefont
  {Thierens}, \citenamefont {D'hondt},\ and\ \citenamefont
  {Deruytter}}]{declercq1536}%
  \BibitemOpen
  \bibfield  {author} {\bibinfo {author} {\bibfnamefont {A.}~\bibnamefont
  {De~Clercq}}, \bibinfo {author} {\bibfnamefont {E.}~\bibnamefont {Jacobs}},
  \bibinfo {author} {\bibfnamefont {D.}~\bibnamefont {De~Frenne}}, \bibinfo
  {author} {\bibfnamefont {H.}~\bibnamefont {Thierens}}, \bibinfo {author}
  {\bibfnamefont {P.}~\bibnamefont {D'hondt}}, \ and\ \bibinfo {author}
  {\bibfnamefont {A.~J.}\ \bibnamefont {Deruytter}},\ }\bibfield  {title}
  {\enquote {\bibinfo {title} {{F}ragment mass and kinetic energy distribution
  for the photofission of $^{235}\mathrm{U}$ and $^{238}\mathrm{U}$ with
  25-{MeV} end-point bremsstrahlung},}\ }\href {\doibase
  10.1103/PhysRevC.13.1536} {\bibfield  {journal} {\bibinfo  {journal} {Phys.
  Rev. C}\ }\textbf {\bibinfo {volume} {13}},\ \bibinfo {pages} {1536--1543}
  (\bibinfo {year} {1976})}\BibitemShut {NoStop}%
\bibitem [{\citenamefont {Vassh}\ \emph {et~al.}(2019)\citenamefont {Vassh},
  \citenamefont {Vogt}, \citenamefont {Surman}, \citenamefont {Randrup},
  \citenamefont {Sprouse}, \citenamefont {Mumpower}, \citenamefont {Jaffke},
  \citenamefont {Shaw}, \citenamefont {Holmbeck}, \citenamefont {Zhu},\ and\
  \citenamefont {McLaughlin}}]{vassh2019}%
  \BibitemOpen
  \bibfield  {author} {\bibinfo {author} {\bibfnamefont {N.}~\bibnamefont
  {Vassh}}, \bibinfo {author} {\bibfnamefont {R.}~\bibnamefont {Vogt}},
  \bibinfo {author} {\bibfnamefont {R.}~\bibnamefont {Surman}}, \bibinfo
  {author} {\bibfnamefont {J.}~\bibnamefont {Randrup}}, \bibinfo {author}
  {\bibfnamefont {T.~M.}\ \bibnamefont {Sprouse}}, \bibinfo {author}
  {\bibfnamefont {M.~R.}\ \bibnamefont {Mumpower}}, \bibinfo {author}
  {\bibfnamefont {P.}~\bibnamefont {Jaffke}}, \bibinfo {author} {\bibfnamefont
  {D.}~\bibnamefont {Shaw}}, \bibinfo {author} {\bibfnamefont {E.~M.}\
  \bibnamefont {Holmbeck}}, \bibinfo {author} {\bibfnamefont {Y.-L.}\
  \bibnamefont {Zhu}}, \ and\ \bibinfo {author} {\bibfnamefont {G.~C.}\
  \bibnamefont {McLaughlin}},\ }\bibfield  {title} {\enquote {\bibinfo {title}
  {Using excitation-energy dependent fission yields to identify key fissioning
  nuclei in r -process nucleosynthesis},}\ }\href {\doibase
  10.1088/1361-6471/ab0bea} {\bibfield  {journal} {\bibinfo  {journal} {J.
  Phys. G}\ }\textbf {\bibinfo {volume} {46}},\ \bibinfo {pages} {065202}
  (\bibinfo {year} {2019})}\BibitemShut {NoStop}%
\bibitem [{\citenamefont {Giuliani}\ \emph {et~al.}(2020)\citenamefont
  {Giuliani}, \citenamefont {Mart{\'{i}}nez-Pinedo}, \citenamefont {Wu},\ and\
  \citenamefont {Robledo}}]{giuliani2019a}%
  \BibitemOpen
  \bibfield  {author} {\bibinfo {author} {\bibfnamefont {S.~A.}\ \bibnamefont
  {Giuliani}}, \bibinfo {author} {\bibfnamefont {G.}~\bibnamefont
  {Mart{\'{i}}nez-Pinedo}}, \bibinfo {author} {\bibfnamefont {M.-R.}\
  \bibnamefont {Wu}}, \ and\ \bibinfo {author} {\bibfnamefont {L.~M.}\
  \bibnamefont {Robledo}},\ }\bibfield  {title} {\enquote {\bibinfo {title}
  {{Fission and the r-process nucleosynthesis of translead nuclei in neutron
  star mergers}},}\ }\href {\doibase 10.1103/PhysRevC.102.045804} {\bibfield
  {journal} {\bibinfo  {journal} {Phys Rev. C}\ }\textbf {\bibinfo {volume}
  {102}},\ \bibinfo {pages} {045804} (\bibinfo {year} {2020})}\BibitemShut
  {NoStop}%
\bibitem [{\citenamefont {Warda}\ \emph {et~al.}(2018)\citenamefont {Warda},
  \citenamefont {Zdeb},\ and\ \citenamefont {Robledo}}]{warda2018}%
  \BibitemOpen
  \bibfield  {author} {\bibinfo {author} {\bibfnamefont {M.}~\bibnamefont
  {Warda}}, \bibinfo {author} {\bibfnamefont {A.}~\bibnamefont {Zdeb}}, \ and\
  \bibinfo {author} {\bibfnamefont {L.~M.}\ \bibnamefont {Robledo}},\
  }\bibfield  {title} {\enquote {\bibinfo {title} {Cluster radioactivity in
  superheavy nuclei},}\ }\href {\doibase 10.1103/PhysRevC.98.041602} {\bibfield
   {journal} {\bibinfo  {journal} {Phys. Rev. C}\ }\textbf {\bibinfo {volume}
  {98}},\ \bibinfo {pages} {041602} (\bibinfo {year} {2018})}\BibitemShut
  {NoStop}%
\end{thebibliography}%
\end{document}